\documentclass[12pt, a4paper]{article}

\usepackage{amsmath}
\usepackage{amsfonts}
\usepackage{amssymb}
\usepackage{graphicx, rotating}
\usepackage{epsfig}
\usepackage{latexsym}
\usepackage{graphicx}
\usepackage{color}
\usepackage{amsmath,bm,amssymb}
\usepackage{cite}
\usepackage{slashed}
\usepackage{hyperref}
\hypersetup{colorlinks, citecolor=bluscuro, linkcolor=black, urlcolor=bluscuro}
\definecolor{rossos}{cmyk}{0,1,1,0.55}
\definecolor{bluscuro}{rgb}{0.15, 0.2, .85}
\definecolor{bluchiaro}{cmyk}{1,.3,0.,0.1}

\newcommand{\sumintK}{\sum_K\mspace{-25mu}\int\;}

\newcommand{\nn}{\nonumber}

\newcommand{\be}{\begin{equation}}
\newcommand{\ee}{\end{equation}}
\newcommand{\bea}{\begin{eqnarray}}
\newcommand{\eea}{\end{eqnarray}}

\newcommand*\DAlambert{\mathop{}\!\mathbin\Box}

 \def\ra {\rightarrow}
\newcommand{\dpsq}{(\partial\phi)^2}
\newcommand{\dpf}{(\partial\phi)^4}
 \newcommand{\dsqp}{(\partial^2\phi)}

\newcommand{\arXiv}[2]{\href{http://arxiv.org/pdf/#1}{{\tt [#2/#1]}}}
\newcommand{\arXivold}[1]{\href{http://arxiv.org/pdf/#1}{{\tt [#1]}}}

 \def\ph{\varphi}

\begin{document}

%FRONTPAGE2%%%%%%
\begin{titlepage}
\begin{flushright}
DESY 16-161\\
CERN-TH-2016-187
\end{flushright}
\vspace{.3in}

\vspace{1cm}
\begin{center}
{\Large\bf\color{black}
Gauge-Independent  Scales Related\\[0.5cm]
 to the Standard Model Vacuum Instability}\\
\bigskip\color{black}
\vspace{1cm}{
{\large J.R.~Espinosa$^{a,b}$, M.~Garny$^c$, T.~Konstandin$^d$, A. Riotto$^e$}
\vspace{0.3cm}
} \\[7mm]
{\it {$^a$\, Institut de F\'isica d'Altes Energies (IFAE), The Barcelona Institute of Science and Technology (BIST), Campus UAB, E-08193, Bellaterra (Barcelona), Spain}}\\
{\it $^b$ {ICREA, Pg. Llu\'is Companys 23, 08010 Barcelona, Spain}}\\
{\it {$^c$\, CERN Theory Department, CH-1211 Geneva, Switzerland}}\\
{\it $^d$ {DESY, Notkestr. 85, 22607 Hamburg, Germany}}\\
{\it $^e$ Department of Theoretical Physics and Center for Astroparticle Physics (CAP)}\\
{\it 24 quai E. Ansermet, CH-1211 Geneva 4, Switzerland}
\end{center}
\bigskip

\vspace{.4cm}

\begin{abstract}
\noindent
{The measured (central) values of the Higgs and top quark masses indicate that  the Standard Model (SM) effective potential develops an instability at high field values. The scale of this instability, determined as the Higgs field value at which the potential drops below the electroweak minimum, is about $10^{11}$~GeV. However, such a scale is unphysical as it is not gauge invariant and suffers from  a gauge-fixing uncertainty of up to two orders of magnitude. Subjecting our system, the SM, to several probes of the instability (adding higher order operators to the potential; letting the vacuum decay through critical bubbles;  heating up the system to very high temperature; inflating it) and asking in each case physical questions, we are able to provide several gauge-invariant scales related with the Higgs potential instability. 
}

\end{abstract}
\bigskip

\end{titlepage}

%%%%%%%%%%%%%%%%%%%%%%%%%%%%%%%%%%%%%%%%
\section{Introduction \label{sec:intro}} 
%%%%%%%%%%%%%%%%%%%%%%%%%%%%%%%%%%%%%%%%

After the LHC runs made so far, with the discovery of a light Higgs boson 
\cite{higgsD} and no signal of additional new physics, we face the possibility that the Standard Model (SM) might describe physics up to very high energy scales, possibly up to the Planck scale. The value of the Higgs mass, measured with great precision by the LHC, which gives the combined value $M_h=125.09\pm 0.21$ (stat) $\pm 0.11$ (syst)  GeV \cite{mhLHC}, turns out to be of particular interest in this context. Large radiative corrections from the heavy top quark
destabilize the Higgs potential at large field values making the electroweak (EW) vacuum metastable. For the current value (from the LHC + Tevatron combination)
$M_t=173.34\pm 0.27$ (stat) $\pm 0.71$ (syst) GeV \cite{mtcomb} we most likely live in such unstable vacuum \cite{stab0}. Intriguingly, we seem to be rather close to the boundary of stability \cite{stab0,stab1,stab2,stab2.5,stab3,stab4} and this translates into a very long lifetime (many orders of magnitude larger than the age of the universe) against decay by quantum tunneling. One concludes that this metastability does not
represent an inconsistency of the SM and cannot be used to argue in favor of new physics. The potential instability has also very interesting cosmological implications \cite{cosmostab0,cosmostab,Zurek,cosmostab2} and might 
have a deeper significance (for some attempts in that direction see \cite{stab1,deep,EFT}).

The instability scale defined as the field value at which the Higgs potential gets lower than the EW vacuum, is quite large, of order $10^{11}$ GeV for the central experimental values of $M_h$ and $M_t$ quoted above.
However, this instability scale turns out to be a gauge-dependent quantity (the previous numerical value corresponds to the potential evaluated, at NNLO, in Landau gauge \cite{stab0,stab1}). 
Such gauge dependence issues \cite{LitVgauge} are well known since the early days of the effective potential \cite{Jackiw} but the problem has attracted some attention recently \cite{DiLuzio,NielsenRedef,AFS} in the wake of the Higgs discovery and the realization that we might be living in a metastable vacuum. The uncertainty in the instability scale due to this gauge dependence was estimated in \cite{DiLuzio} to be potentially sizeable, of up to  two orders
of magnitude. 
 
The goal of this paper is to address this issue by deriving physical gauge-independent scales associated to the instability scale (with varying degrees of how direct the connection is). Following previous discussions and to ease the comparison with earlier literature we use Fermi (or Lorentz) gauge, using the gauge fixing parameter $\xi$ to track the gauge dependence of our results.  The gauge dependence of the effective potential (or more fundamentally of the effective action) is described by the so-called Nielsen identity, which we review in Section~\ref{sec:NI}. After showing explicitly
the gauge dependence of the instability scale in Fermi gauge in Section~\ref{sec:hIg}
we then discuss several ways of extracting physical scales associated with the potential instability. This instability can be cured
by heavy physics that affects the potential through non-renormalizable operators, say $\lambda_6 |H|^6/\Lambda^2$, where $\Lambda$
is the mass scale characterizing the heavy physics. Even though the
potential is a gauge-dependent object, the mass scale $\Lambda$ required
to stabilize the potential turns out to be gauge-independent. We discuss this particular proposal (exploring also  non-renormalizable operators of higher orders) in Section \ref{sec:NRO}. We show how in this way the instability scale can be determined unambiguosly
and lies within one order of magnitude of the naive instability scale calculated in Landau gauge. 

A different scale related to the instability can be obtained from the radius $R_c$ of the critical bubble for vacuum decay. We discuss this in Section~\ref{sec:Rc}, giving a gauge-independent definition of this radius, including also gravity effects. The energy scale associated to this critical radius is the scale at which new physics can have a direct impact on the vacuum lifetime. In the case of the SM it is much heavier that  the instability scale itself, being rather close to the Planck scale. 

In Section~\ref{sec:Tc} we also study the behaviour of the unstable potential at very high temperatures and obtain a critical temperature at which there is a degeneracy between the EW minimum and the one at very high field values. This temperature, which can be proven to be gauge invariant, turns out to be too loosely related to the instability scale in the SM and not too illuminating. 

Finally, in Section \ref{sec:HI} we discuss how to probe the instability scale via inflation, that causes fluctuations in the Higgs field proportional to the Hubble rate
$H_I$  and makes it probe the unstable region if $H_I$ is large  (comparable to the instability scale). We prove that the probability of finding the Higgs in a certain field range, after a given number of e-folds of inflation, is a gauge-invariant quantity. Then we discuss how to extract a value for the Hubble rate that reflects closely the scale of the potential instability with the advantage of being a gauge-invariant quantity.

After drawing some conclusions, we collect some technical details and results in several Appendices.  In Appendix~\ref{app:RGES} we calculate the renormalization group equations for the Wilson coefficients of higher order operators added to the effective potential.  In Appendix~\ref{app:derexp} we derive the $\xi$-dependence of the different functions that appear in the effective action (and the energy-momentum tensor derived from it) when using a derivative expansion, up to ${\cal O}(\partial^4)$. In Appendix~\ref{app:NIT} we discuss the validity of the Nielsen identity at finite temperature, deriving explicit results for the SM at one-loop.

%%%%%%%%%%%%%%%%%%%%%%%%%%%%%%%%%%%%%%%%
\section{The Nielsen Identities \label{sec:NI}}
%%%%%%%%%%%%%%%%%%%%%%%%%%%%%%%%%%%%%%%%

The dependence of the Higgs effective potential on the gauge-fixing parameters $\xi$ derives from the gauge dependence of the effective action $S$ itself. Nevertheless, the potential and the effective action are very useful and it is possible to extract from them physical quantities that are gauge-independent. 

The Nielsen identity \cite{Nielsen,FukudaKugo,Aitchison} describes the $\xi$-dependence of the effective action and plays a  central role in discussing how to obtain gauge-independent quantities. For cases with a Higgs background only, the identity reads
\be
\label{NIS}
\xi \frac{\partial }{\partial\xi}S[h(x),\xi]=-\int d^4 y \ K[h(y)] \frac{\delta S}{\delta h(y)}\ ,
\ee
where $K[h(y)]$ is a known functional of $h$, given in~\cite{Nielsen}. According to this identity, the effective action evaluated on a solution of the equation of motion (EoM) for $h$, (that is, $\delta S/\delta h=0$) is $\xi$-independent. 
A particular instance of this general result is the $\xi$-independence of the values of the effective potential at its extrema, a well known result.

If one writes the effective action in a derivative expansion
\be
S[h] = \int  d^4 x \left[-V(h)+ \frac{1}{2}Z(h)(\partial_\mu h)^2 + {\cal O}(\partial^4)\right] \ ,
\label{Sder}
\ee
a series of Nielsen identities  for the coefficient-functions in this expansion can be derived from the identity in Eq.~(\ref{NIS}). We show this explicitly, up to fourth order in the expansion, in Appendix~\ref{app:derexp}. At the lowest order in this derivative expansion, {\em i.e.} for constant field configurations, one finds 
a Nielsen identity for  the effective potential:
\be\label{NIV}
\xi\frac{\partial V}{\partial  \xi} + C(h)  V'=0
\ ,
\ee
where $C(h)$ is the value of $K[h]$ for constant $h$. As anticipated above, this identity implies the $\xi$-independence of the values of the potential at the extremal points. 
The identity (\ref{NIV}) can be rewritten as 
\be
\xi \frac{dV}{d\xi}= \xi \frac{\partial V}{\partial\xi} + \frac{\partial V}{\partial h}\,\xi\frac{\partial h}{\partial\xi}=0\ ,
\ee 
showing that the explicit $\xi$-dependence of the effective potential  can be compensated by an implicit $\xi$-dependence of the field as:
\be\label{NIh}
\xi \frac{d h}{d  \xi} = C(h)\ .
\ee
A change in $\xi$ is therefore equivalent to the field redefinition
(\ref{NIh}). This way of looking a the effect of a change in $\xi$ makes it obvious that the values of the potential at the extremal points are gauge-independent.

Next, we derive Nielsen identities for the equation of motion and for the energy-momentum tensor $T^{\mu\nu}$  that allow us to understand the 
$\xi$-dependence of the solutions of the EoM and prove the $\xi$-independence of $T^{\mu\nu}$, both of which are needed for later discussions. We begin with the Nielsen identity for the effective action given in Eq.~(\ref{NIS}).
To derive a Nielsen identity for the equation of motion, we take a functional derivative with respect to $h(x)$, getting
\be
  \xi \frac{\partial}{\partial\xi}\frac{\delta S}{\delta h(x)} + \int d^4y \left( \frac{\delta^2 S}{\delta h(y)\delta h(x)} K[h(y)] + \frac{\delta S}{\delta h(y)} \frac{\delta K[h(y)]}{\delta h(x)}\right) = 0\;.
\ee
Evaluating the effective action for a field $\bar h(x, \xi)$ 
that solves the EoM,  $\delta S/\delta h=0$, the last term 
in the  equation above drops, while the first two terms  can be combined into a total derivative,
\be
  \xi \frac{d}{d\xi}\frac{\delta S[\bar h,\xi]}{\delta h(x)} = 0\; ,
\label{dGdxi}
\ee
provided $\bar h$ fulfills 
\be\label{phibar}
  \xi \frac{d}{d\xi}\bar h(x,\xi) = K[\bar h(x)]\, .
\ee
As the total derivative (\ref{dGdxi}) is zero, this in fact shows that the solution of the EoM for $\xi+d\xi$ is $\bar h + K[\bar h] d\log\xi$.
In other words, the solution $\bar h(x,\xi)$ of (\ref{phibar}) 
describes  how a solution of the EoM changes when varying the gauge parameter $\xi$.\footnote{It is worth noting that the value of the action for any field configuration, even off-shell, is invariant under the combined $\xi$- and  field change implied by (\ref{phibar}), as is evident from Eq. (\ref{NIS}).}

Let us next discuss the $\xi$-independence of the energy-momentum tensor, that is defined through the variation of the 
action under changes of the metric tensor, as
\be
  T^{\mu\nu}(x) = 
   -\frac{2}{\sqrt{-g}}\frac{\delta S}{\delta g_{\mu\nu}(x)}\ ,
\ee
where $g$ is the determinant of the metric.  We are ultimately interested in a flat background, but keep the metric explicit in order
to derive the energy-momentum tensor.  In order to derive a Nielsen identity for the energy-momentum tensor\footnote{
 This requires that the Nielsen identity not only holds in flat Minkowski space but for an arbitrary background metric (at least in the neighbourhood of flat space). Such generalized Nielsen identity can be obtained by writing all terms in a general coordinate invariant form. Gauge-fixing of gravity is not required for a non-dynamical background metric which is all we need here to define the energy-momentum tensor. The usual proof of the Nielsen identity is then directly carried over to the non-flat case. To calculate explicitly $K[h,g]$ is nevertheless much more cumbersome. } we take a derivative of (\ref{NIS}) with respect to $g_{\mu\nu}(x)$,
\be
  \xi \frac{\partial}{\partial\xi}T^{\mu\nu}(x) - \frac{2}{\sqrt{-g(x)}} \int d^4y \left( \frac{\delta S}{\delta h(y)\delta g_{\mu\nu}(x)} K[h(y)] + \frac{\delta S}{\delta h(y)} \frac{\delta K[h(y)]}{\delta g_{\mu\nu}(x)} \right) = 0\;.
\ee
When evaluated for $\bar h(x,\xi)$, a solution of the EoM, the first two terms can again be combined into a total derivative, while the last term vanishes, and we get
\be
  \xi \frac{d}{d\xi}T^{\mu\nu}(x)\Big|_{h=\bar h(x,\xi)}= 0\;.
\ee
This means that the explicit gauge parameter dependence is
precisely compensated by the change of the field value when varying $\xi$. Therefore, the total $\xi$-dependence vanishes, such that on-shell the energy-momentum tensor is gauge-fixing independent.

For the particular case in which we are interested, the SM in Fermi gauge, we have in fact two $\xi$ parameters appearing in the 
EW gauge-fixing Lagrangian:
\be
{\cal L}_{gf} = -\frac{1}{2\xi_B} (\partial^\mu B_\mu)^2
-\frac{1}{2\xi_W} (\partial^\mu W^a_\mu)^2\ ,
\ee
and the effective potential depends on both of them. 
We have a Nielsen identity for each, with 
\be\label{NIVs}
\xi_i\frac{\partial V}{\partial  \xi_i} + C_i(h)  V'=0
\ ,
\ee
for $\xi_i=\xi_B,\xi_W$, with the $C_i(h)$ functions given by
\bea
\label{CB}C_B(h)&=&\frac{i g'}{2}\int d^4 y\ \langle
c(x) \chi^0(x) \bar c(y) \partial_\mu B^\mu(y)
\rangle\ , \\
\label{CW}C_W(h)&=&\frac{i g}{2}\int d^4 y\ \langle
c_a(x) \chi_a(x) \bar c_b(y) \partial_\mu W_b^\mu(y)
\rangle\ ,
\eea
where $c,\bar c$ ($c_a,\bar c_a$) are the $U(1)_Y$ [$SU(2)_L$] ghost fields and $\chi_a$ the Goldstone boson fields
while $B_\mu$ and $W_\mu^a$ are the $U(1)_Y$ and $SU(2)_L$ gauge bosons, respectively and $h$ is a constant background.

Let us write the tree-level potential as
\be
\label{V0}
V_0(h)=-\frac12 m^2 h^2 +\frac14 \lambda h^4 \ .
\ee 
The renormalized  one-loop potential (in $\overline{\rm MS}$ scheme) is obtained as \cite{CW}
\be
\label{V1}
V_1(\phi)=\sum_\alpha N_\alpha J^0_{\alpha}(M_\alpha^2)  =\frac{\kappa}{4}\sum_\alpha N_\alpha M_\alpha^4
\left[\log\frac{M_\alpha^2}{\mu^2}-C_\alpha \right]\ ,
\ee
where $\kappa=1/(16\pi^2)$, $\mu$ is the renormalization scale
and the index $\alpha$ runs over different particle species, with  $N_\alpha$ degrees of freedom (taken negative for fermions). 
The squared masses $M_\alpha^2$ are the corresponding masses in an $h$ background. 
The main contributions to the potential come from:
\be
\label{spectrum}
\hspace{-0.5cm}
\begin{array}{rll}
{\rm Top\ quarks:} &  N_t=-12\ , &  M_t^2 = \frac12 y_t^2 h^2\ , \\
W^\pm \ {\rm bosons} : & N_W=6\ , &  M_W^2 = \frac14 g^2 h^2\ , \\
Z^0\  {\rm bosons }: & N_Z=3\ , &   M_Z^2 = \frac14 (g^2+{g'}^2) h^2\ , \\
{\rm Higgs\ bosons } : & N_h=1\ , &  M_h^2 = -m^2 + 3\lambda h^2\ , \\
{\rm Neutral\ Goldstones} : & N_{B_\pm}=1\ , &  M_{B_\pm}^2 = \frac12[M_G^2\pm\sqrt{M_G^4-4(\xi_W M_W^2+\xi_B M_B^2)M_G^2}]\ , \\
{\rm Charged\ Goldstones} : & N_{A_\pm}=2\ , &  M_{A_\pm}^2 = \frac12(M_G^2\pm\sqrt{M_G^4-4\xi_W M_G^2 M_W^2})\ , 
\end{array}
\ee
where we use the auxiliary squared masses
\be
M_B^2\equiv \frac14 {g'}^2h^2 \ ,\quad M_G^2\equiv \frac{1}{h}\frac{\partial V_0}{\partial h}=
-m^2 + \lambda h^2\ .
\ee
Finally, the $C_\alpha$s are constants: $C_\alpha=3/2$ for scalars and fermions and $C_\alpha=5/6$ for gauge bosons.

Fermi gauge is afflicted  by infrared divergences beyond those generic in all gauges (with massless Goldstone bosons) that remain even after the resummation cure
that fixes the latter \cite{Martin,EEK}. For a discussion of this issue (more precisely an IR divergence in the first derivative of the effective potential) and its solution(s), see  \cite{EGK}. In order to avoid this complication we use a Fukuda-Kugo IR
regulator $\mu_{IR}$ as described in Subsection~4.2 of \cite{EGK}. In this case one needs to add to the potential also ghost contributions with $N_{c_Z}=-2, N_{c_W}=-4, M_{c_Z}^2=\mu_{IR} M_Z, M_{c_W}^2=\mu_{IR} M_W, C_{c_i}=3/2$. We  checked that none of our results depends on the choice of $\mu_{IR}$.

The functions $C_{B,W}(h)$ that enter in the Nielsen identity in Eq.~(\ref{NIV}), calculated at one loop, are
\bea
\label{CB0}
C^{(1)}_{B}(h)&=&\frac{\xi_B}{8} {g'}^2 h\left[\frac{I^0_{B_+}-I^0_{B_-}}{M^2_{B_+}-M^2_{B_-}}\right]
\ , \\
\label{CW0}
C^{(1)}_{W}(h)&=&\frac{\xi_W}{8} {g}^2 h
\left[\frac{I^0_{B_+}-I^0_{B_-}}{M^2_{B_+}-M^2_{B_-}}+2 
\frac{I^0_{A_+}-I^0_{A_-}}{M^2_{A_+}-M^2_{A_-}}
\right]\ ,
\eea
with 
\be
I^0_\alpha(M_\alpha^2)\equiv  2\frac{\partial}{\partial M_\alpha^2}J^0_\alpha(M_\alpha^2)= \kappa M_\alpha^2\left(\log\frac{M_\alpha^2}{\mu^2}-C_\alpha+\frac12\right)\ .
\ee
With the previous expressions it is  straightforward to check that the one-loop Nielsen identities
\be
\xi_i \frac{\partial V_1(h)}{\partial \xi_i} + C^{(1)}_i \frac{\partial V_0(h)}{\partial h}=0\ ,
\ee
are indeed fulfilled. For the numerical analyses in the following sections we set $\xi_B=\xi_W=\xi$ at the EW scale (more precisely at $\mu=M_t$) as was
done in \cite{DiLuzio}, and discuss the dependence of different quantities with $\xi$.

%%%%%%%%%%%%%%%%%%%%%%%%%%%%%%%%%%%%%%%%
\section{Gauge Dependence of the Instability Scale \label{sec:hIg}}
%%%%%%%%%%%%%%%%%%%%%%%%%%%%%%%%%%%%%%%%

\begin{figure}[t]
\begin{center}
  \includegraphics[width=0.8\textwidth]{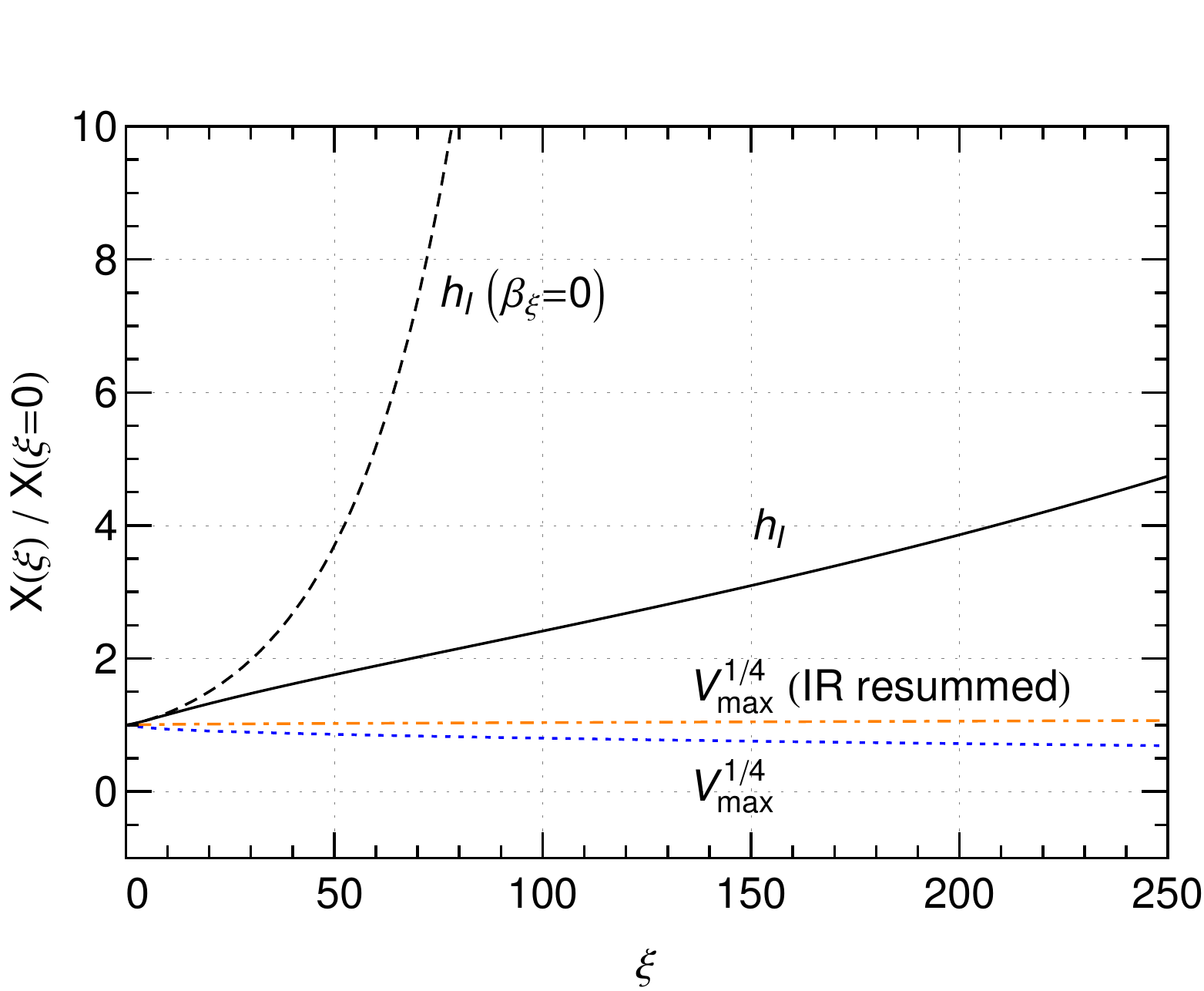}
\end{center}
\caption{\label{fig:instScale2}%
\em Gauge dependence of several quantities normalized to their value for $\xi=0$. Black lines give the scale $h_I(\xi)$ at which the effective potential drops below the value of the electroweak minimum taking into account the running of the gauge fixing parameters (solid) or keeping them fixed (dashed). For comparison, we also show the value of the effective potential at the maximum, $V_{max}^{1/4}$, with or without resummation of IR divergences, as indicated.
}
\end{figure}

As discussed in the Introduction, for the measured values of the Higgs and top quark masses, the Standard Model develops an instability at high field values. One can take as the scale of the instability the field value $h_I$ at which the effective potential drops below the value of the electroweak minimum. Given the order of magnitude of the scales involved this corresponds in practice to $V(h_I)=0$, with $h_I\simeq 10^{11}$ GeV for the central values of $M_h$ and $M_t$. It is well-known that $h_I$
is not a gauge-independent quantity and it was calculated recently,
using Fermi and $R_\xi$ gauges \cite{DiLuzio}, that the gauge dependence leads to an uncertainty of up to two orders of magnitude in $h_I$ (if one takes extreme values $\xi\sim 300$, at the limit of validity of the perturbative regime).
For illustration, the gauge dependence of $h_I$ in Fermi gauge is shown in Fig.\,\ref{fig:instScale2}. The black solid line corresponds to the result extracted from the one-loop RG improved effective potential, taking into account also the running of the gauge fixing parameters. It turns out that even when fixing $\xi_B=\xi_W=\xi$ to large values at the EW scale, the running significantly reduces their value in the UV \cite{DiLuzio} reducing the impact of gauge dependence.  This is illustrated by the black dashed line, for which the running of the gauge-fixing parameters has been neglected.  These two lines reproduce the results of \cite{DiLuzio}.

One might ask whether the gauge dependence, especially for very large values of $\xi$, signals a poor perturbative description rather than the expected gauge dependence of $h_I$. To address this question we also show the value of the potential evaluated at the maximum in Fig.\,\ref{fig:instScale2}, which is a gauge-independent quantity. Its residual gauge dependence due to the perturbative computation of the potential is indeed small (blue dotted line).  This dependence is further reduced to the few $\%$ level for the effective potential with IR resummed Goldstone mass parameter \cite{Martin,EEK} (orange dot-dashed line) in agreement with expectations \cite{EGK}. In addition, we checked that the gauge dependence of $h_I$ agrees,  at the same level of accuracy, with the one expected from the Nielsen identity, obtained by solving the differential equation Eq.\,(\ref{phibar}).
Note that the gauge dependence of $h_I$ in $R_\xi$ gauges is of similar magnitude, but goes in the opposite direction \cite{DiLuzio}, such that the value of $h_I$ varies over a large range.

One pragmatic attitude concerning the gauge dependence of $h_I$ 
is simply to ignore the problem altogether: an order of magnitude estimate of the instability scale, especially when it is so high, might be good enough and the Landau gauge calculation should give a quite reasonable estimate of that scale (why would one take such large values of the gauge fixing parameter as $\xi\sim300$?). Moreover,  the same problem affects the vacuum expectation value  of the Higgs in the electroweak vacuum. That is, the usual $v_{EW}=246$ GeV is also a gauge dependent quantity.
However, as long as one calculates physical quantities the gauge dependence should drop out. Residual gauge dependence might still be left over from truncating perturbative expansions, in which case taking large values of the gauge fixing parameters would be ill advised. Nevertheless, as a point of principle, it is interesting to think of what physical scales are there that can be associated to the presence of the instability of the potential. We discuss several possibilities in the following Sections.

%%%%%%%%%%%%%%%%%%%%%%%%%%%%%%%%%%%%%%%%
\section{Stabilization by Irrelevant Operators \label{sec:NRO}}
%%%%%%%%%%%%%%%%%%%%%%%%%%%%%%%%%%%%%%%%

In this section we explore one physical scale, independent of the gauge-fixing parameter, that reflects in a very clean way the underlying instability scale in the SM effective potential. The basic idea is to probe the behaviour of the theory at high field values by adding a fictitious higher-dimensional operator
to the Higgs potential, characterized by a suppression scale 
$\Lambda$.\footnote{ Generically, if there is really new physics at the scale $\Lambda$ the use of an effective theory to study the impact on the heavy physics on the instability scale, requires a wide separation between the two scales,  
see \cite{BdCE} for a discussion of this point. See also \cite{Eichhorn:2015kea} for a recent analysis (based on functional RG methods) of the potential instability in the presence of higher-dimensional operators.}
At tree-level the potential is
\be
  V_0(h) = -\frac12 m^2 h^2 + \frac{\lambda}{4}h^4 + c_n \frac{h^n}{2^{n/2}\Lambda^{n-4}}\,,
\ee
and we consider only even powers of $n$, with $n\geq 6$, as the non-renormalizable
operator should arise from powers of $(H^\dagger H)$.

\begin{figure}[t]
\begin{center}
 \includegraphics[width=0.4\textwidth]{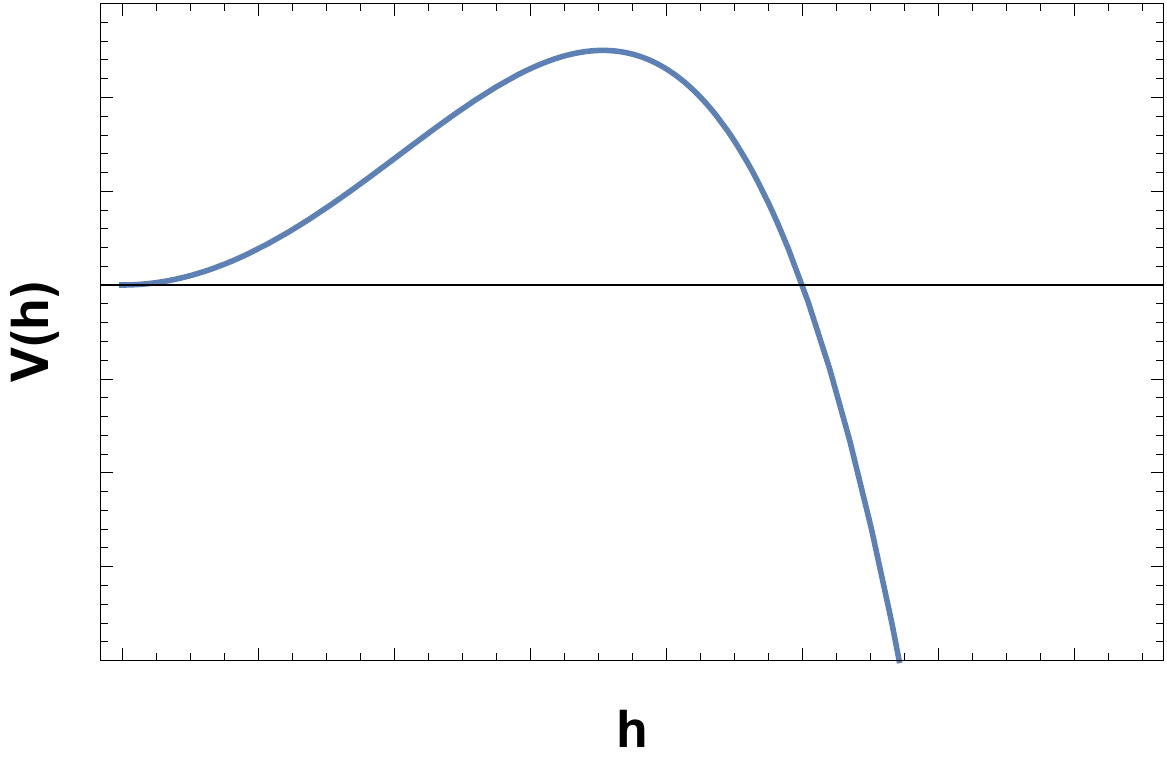}
 \includegraphics[width=0.1\textwidth,height=0.2\textwidth]{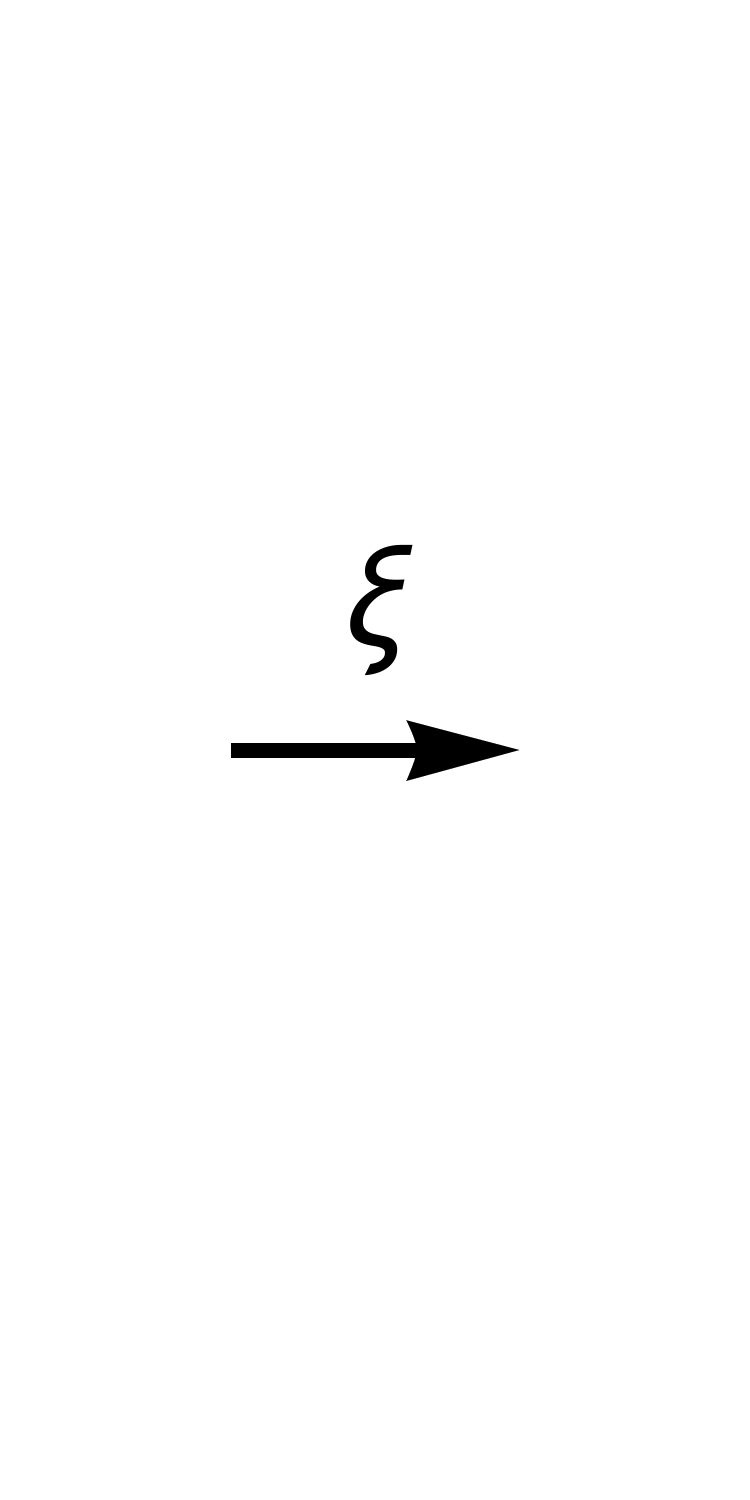}
 \includegraphics[width=0.4\textwidth]{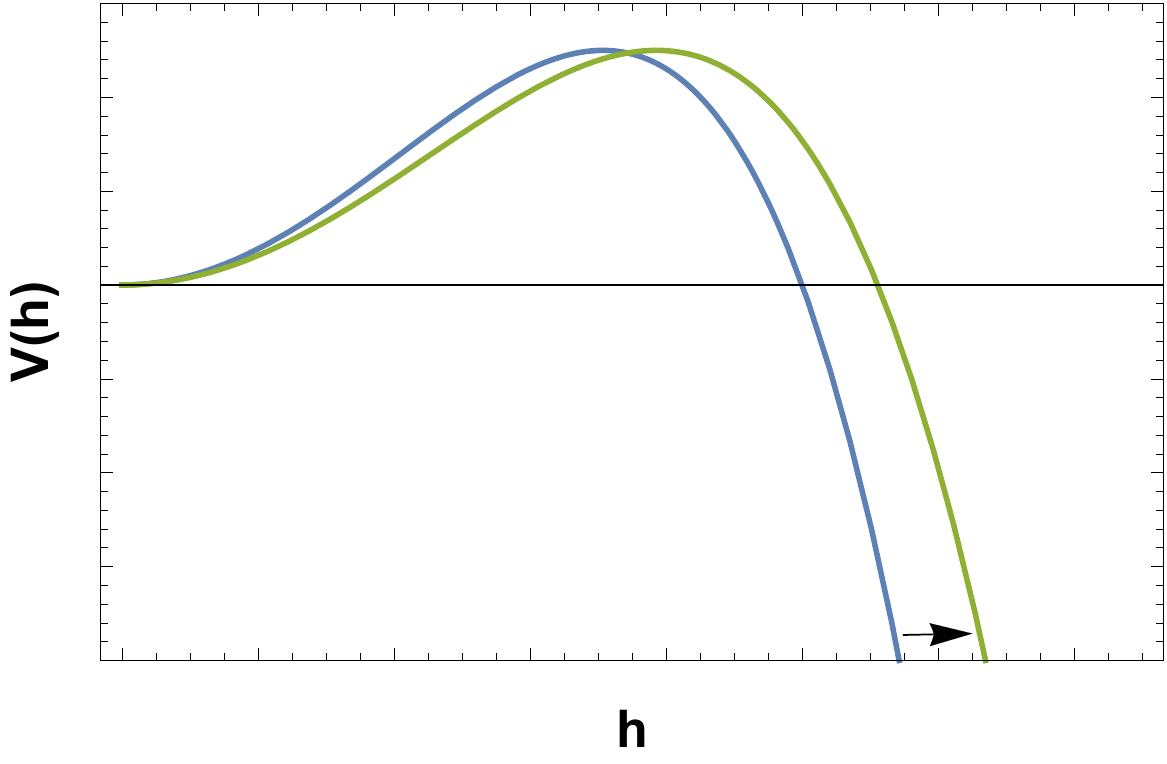}\\
 \includegraphics[width=0.2\textwidth]{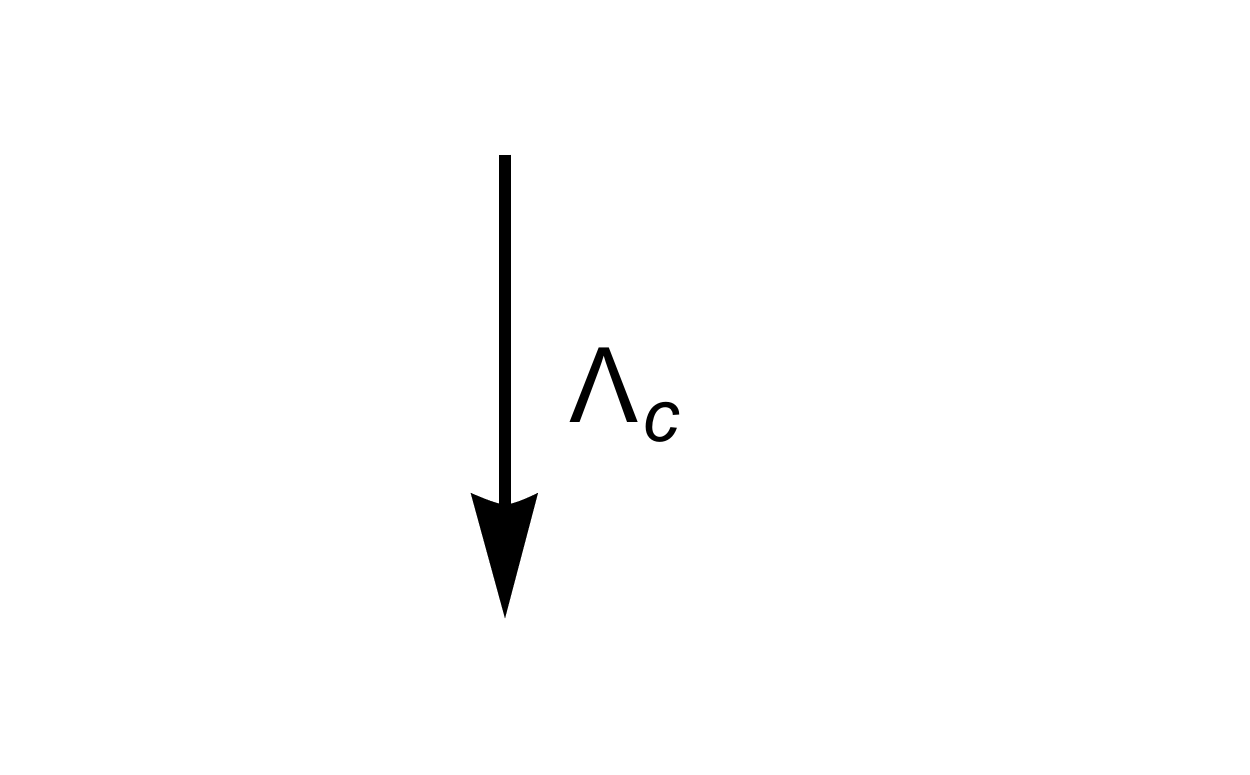}\hspace{4cm}
 \includegraphics[width=0.2\textwidth]{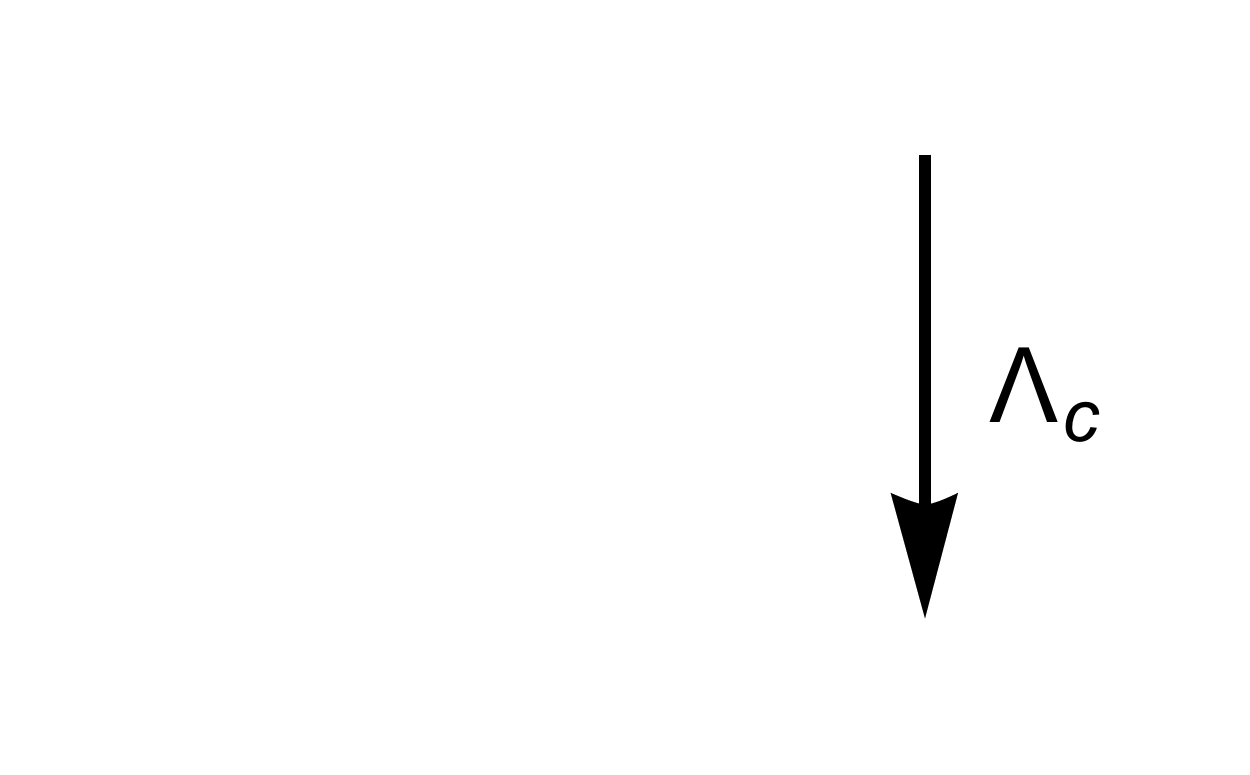}\\
 \includegraphics[width=0.4\textwidth]{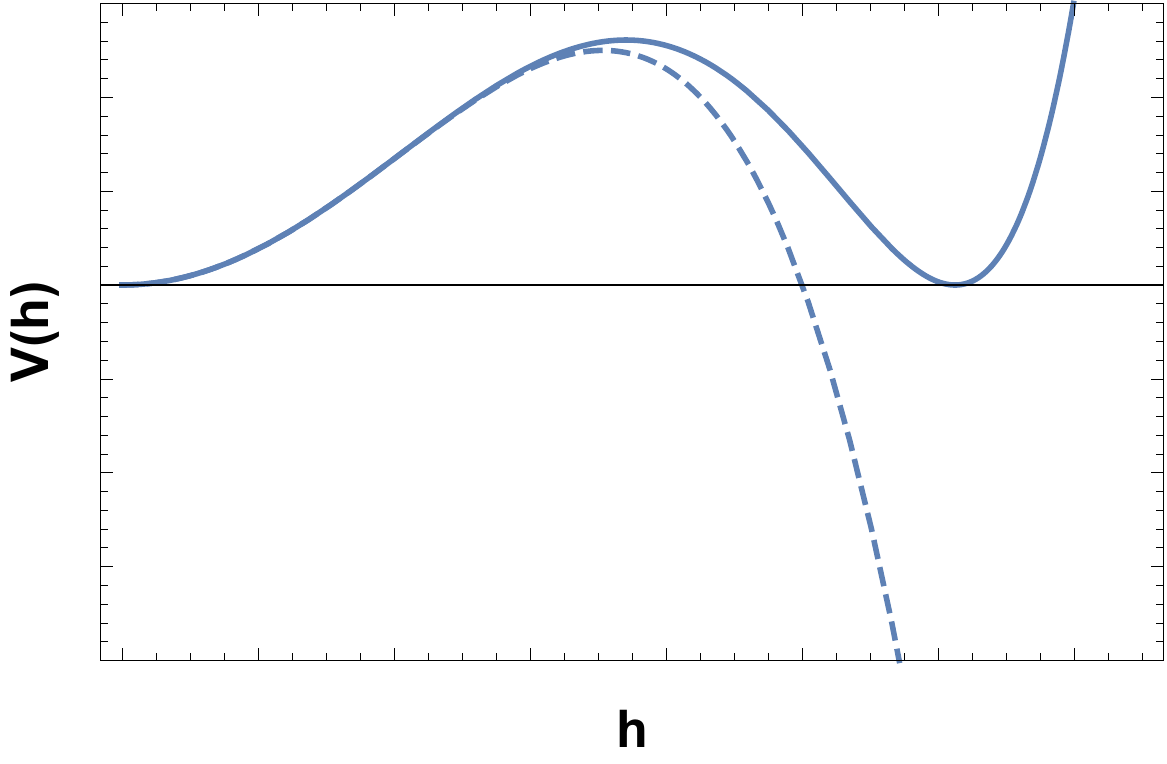}
 \includegraphics[width=0.1\textwidth,height=0.2\textwidth]{p12}
 \includegraphics[width=0.4\textwidth]{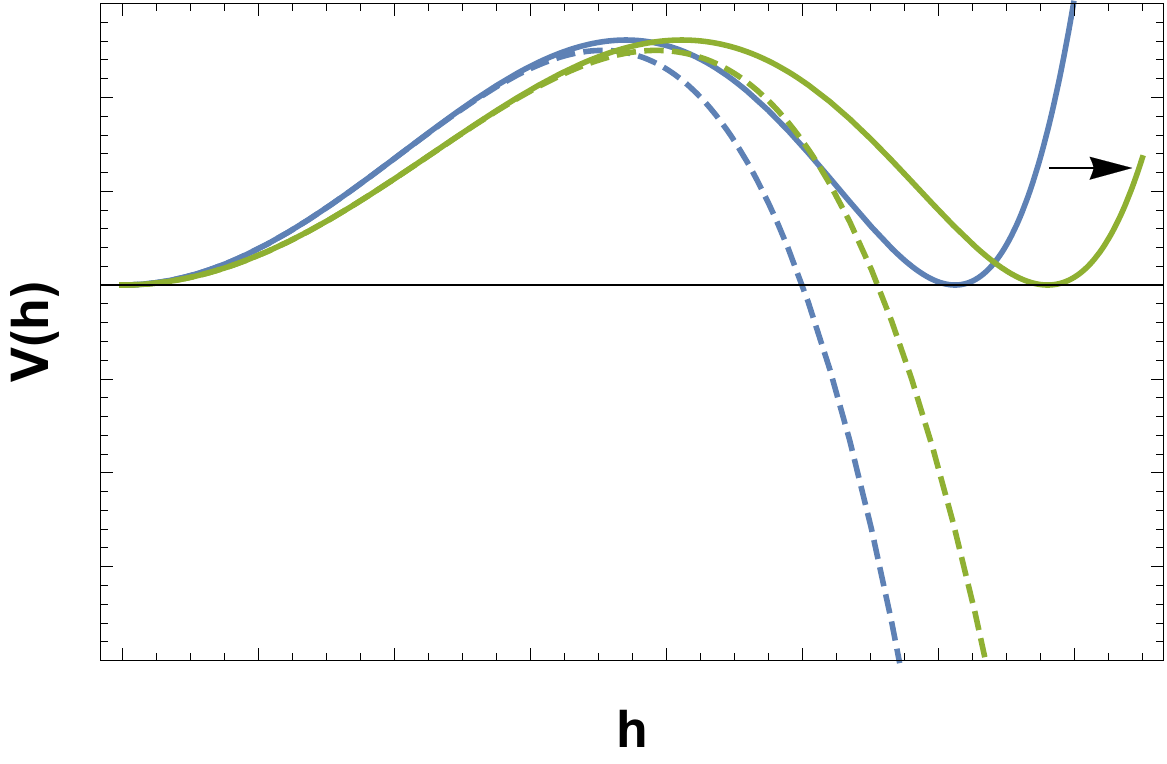}
\end{center}
\caption{\label{fig:NRO}%
\em Illustration of the $\xi$-independence of the scale $\Lambda_c$
defined in the text. Upper left panel: $V(h,\xi)$. Upper right panel: $V(h,\xi+\Delta\xi)$ in green, shows the rescaling effect of changing $\xi\to\xi+\Delta\xi$. Lower left panel: $V(h,\xi;\Lambda_c)$ shows the effect of the non-renormalizable term suppressed by $\Lambda_c$ that gives two degenerate minima. Lower right panel: shows in green the effect of $\xi$ change on previous potential. The degenerate minima stay degenerate.
}
\end{figure}

After including radiative corrections the SM part of the potential develops the instability at some scale [at LO this is captured by $\lambda(h)$ turning negative] but, for positive $c_n$, the new term
stabilizes the effective potential at even larger field values. 
By varying $\Lambda$ (keeping $c_n$ fixed) it is possible to find a critical value of 
$\Lambda_{c}$ for which the high-scale minimum at $h_c$ [$ V'(h_c,\xi;\Lambda_c)= 0$] and the EW minimum at $v_{EW}$ are degenerate,
\be
  V(h_c,\xi;\Lambda_c) = V(v_{EW},\xi;\Lambda_c) \approx 0\,.
\ee
Since the shape of the effective potential depends on the particular value of the gauge-fixing parameter $\xi$ (as indicated) one might expect that the critical value $\Lambda_c$ required to fullfill the conditions above would also be $\xi$-dependent. Interestigly
this is not the case and $\Lambda_c$ is a gauge-independent scale.
This follows from the fact that the value of the potential at an extremal point is independent of $\xi$. The argument is illustrated
by Fig.~\ref{fig:NRO}: a change in $\xi$ affects the unstable potential $V(h,\xi)$ shown in the upper left panel as a field 
redefinition
and $V(h,\xi+\Delta\xi)$ is shown in green in the upper right corner. 
The lower left panel shows instead the effect of adding to the
original potential a non-renormalizable term adjusted to the critical value $V(h,\xi;\Lambda_c)$, such that the potential has two degenerate minima. It is then obvious that the $\xi$ transformation of this potential, $V(h,\xi+\Delta\xi;\Lambda_c)$,
still has two degenerate minima and therefore $\Lambda_c$ is also the critical scale for the potential $V(h,\xi+\Delta\xi)$ and one concludes that $\Lambda_c$ is $\xi$-independent.
On the other hand, the field value $h_c$ at which the high scale minima appears does depend on the gauge fixing. This suggests the use of $\Lambda_c$ instead of $h_I$ as a physical measure of the instability scale.

We describe below how to implement this idea in practice. The effect of the higher-dimensional operator in the one-loop effective potential is straightforward to take into account, simply
using the modified Higgs and Goldstone field-dependent squared masses 
\bea
M_H^2(h) &=& V_0''(h) = -m^2 + 3\lambda h^2 +  \frac{n(n-1)c_n h^{n-2}}{2^{n/2}\Lambda^{n-4}} \ ,\nn\\ 
M_G^2(h) &=& \frac{1}{h}V_0'(h)=-m^2 + \lambda h^2 +  \frac{nc_nh^{n-2}}{2^{n/2}\Lambda^{n-4}}\ ,
\label{MHMG}
\eea
where primes denote $h$-derivatives.
In addition, we take into account IR resummation of the Goldstone mass, as discussed in \cite{EGK}, to minimize residual gauge dependences.  At the order we work, this amounts to adding to $M_G^2$
the one-loop Goldstone self-energy $\Pi_g$ given by
\be
\frac{1}{\kappa}\Pi_g=\frac{dM_H^2}{dh^2} M_H^2(L_H-1)-6y_t^2 M_t^2(L_t-1)+\frac32 g^2 M_W^2 \left(L_W-\frac13\right)+\frac34 (g^2+{g'}^2)M_Z^2\left(L_Z-\frac13\right)\ ,
\ee
where $L_X=\log(M_X^2/\mu^2)$.

The renormalization group (RG) improved effective potential
resums large logarithms between the electroweak and high-scale
minima  in the usual way (see e.g. \cite{stab0}).
For our purpose it is enough to use the one-loop potential  (with IR resummed Goldstone mass)
with SM parameters running at two loops (the renormalization group equations in Fermi gauge can be found in \cite{DiLuzio}) while we treat the effect of
the non-renormalizable operator at leading order, with $c_n$ running at one loop.
The one-loop beta function for $c_n$, $\beta_{c_n}\equiv dc_n/d\log{\mu}$, neglecting corrections of order $m^2/\Lambda^2\ll 1$, is given by (see Appendix~\ref{app:RGES})
\be
\label{eq:betan}
  \beta_{c_n}= 3\kappa\left(n\lambda +y_t^2 -\frac{{g'}^2}{4} -\frac{3g^2}{4}\right) n c_n 
 + \frac{\kappa}{8} \sum_{m=6}^{n-2} m\ p(n,m)\left[m\ p(n,m)+n\right]c_{m}c_{p(n,m)}\ ,
\ee
where $p(n,m)\equiv n+4-m$.
Once we introduce the higher order operator in the SM potential the theory is not renormalizable and further operators of even higher order are generated radiatively. This is reflected in the RG equation above, through the $c_{m}c_{p(n,m)}$ terms. 
We impose the boundary condition $c_n(\Lambda)=1$, $c_{m\not= n}(\Lambda)=0$. Close to the scale $\Lambda$ the operator $h^n$ dominates and we can safely neglect the rest for our analysis.

The non-renormalizable couplings are completely irrelevant at the electroweak scale. For the renormalizable couplings we impose boundary conditions at the electroweak scale, more precisely at $\mu_0\equiv M_t=173.34\,$GeV,
using the best-fit values given in \cite{stab1}.  We first consider the lowest choice $n=6$, and then discuss what happens for higher $n$.

\begin{figure}[!t]
\begin{center}
  \includegraphics[width=0.45\textwidth]{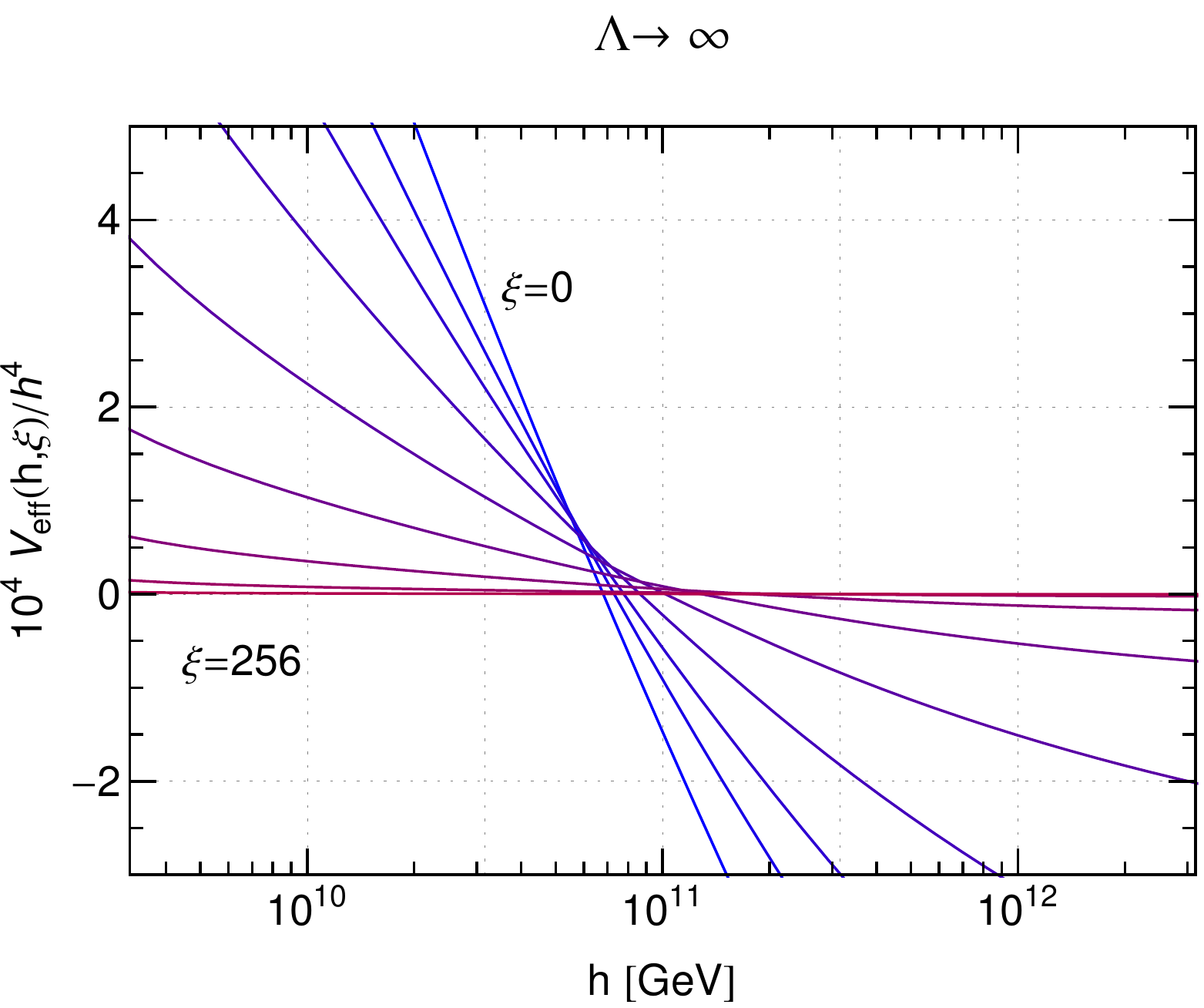}
  \includegraphics[width=0.45\textwidth]{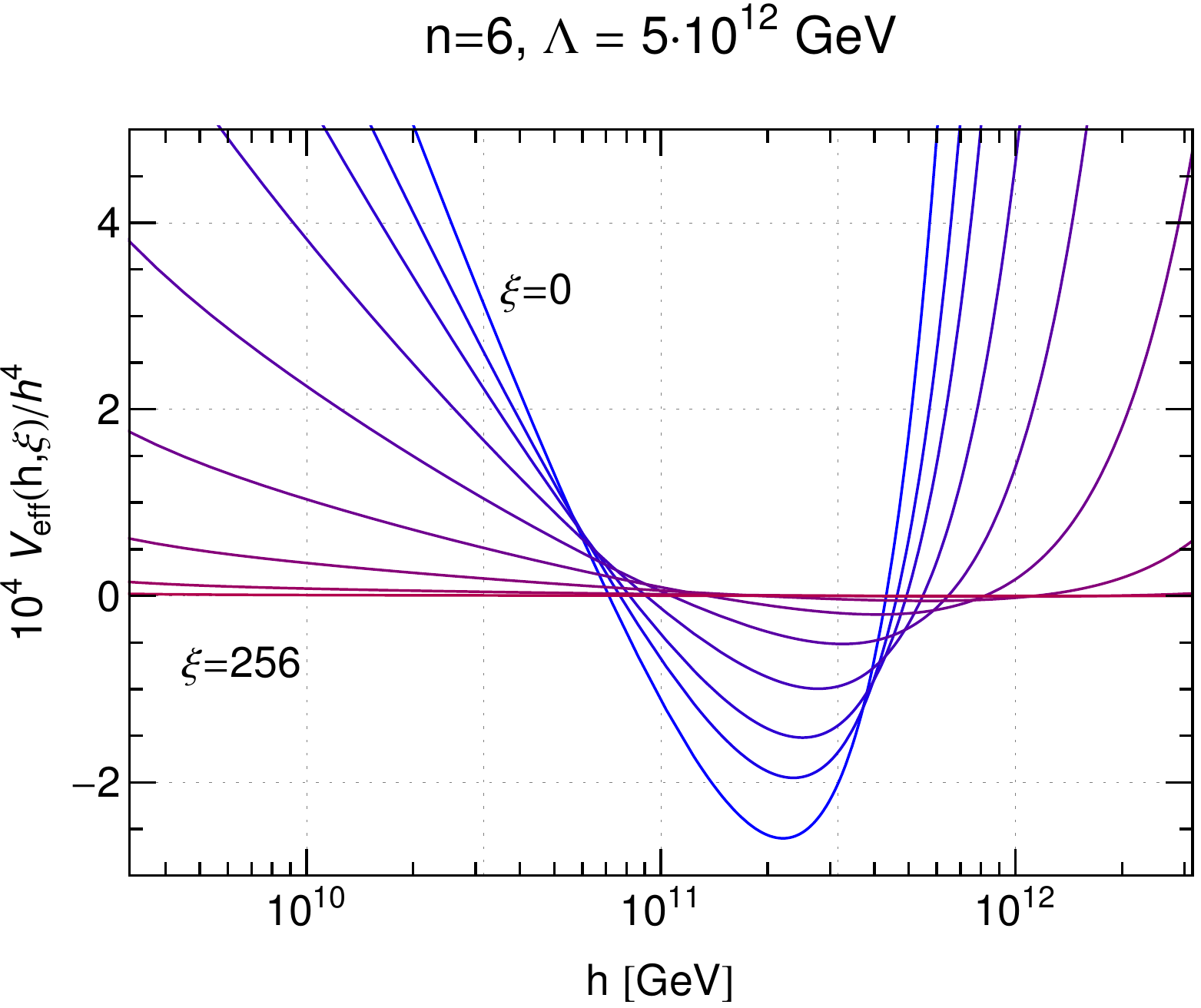}\\[2ex]
  \includegraphics[width=0.45\textwidth]{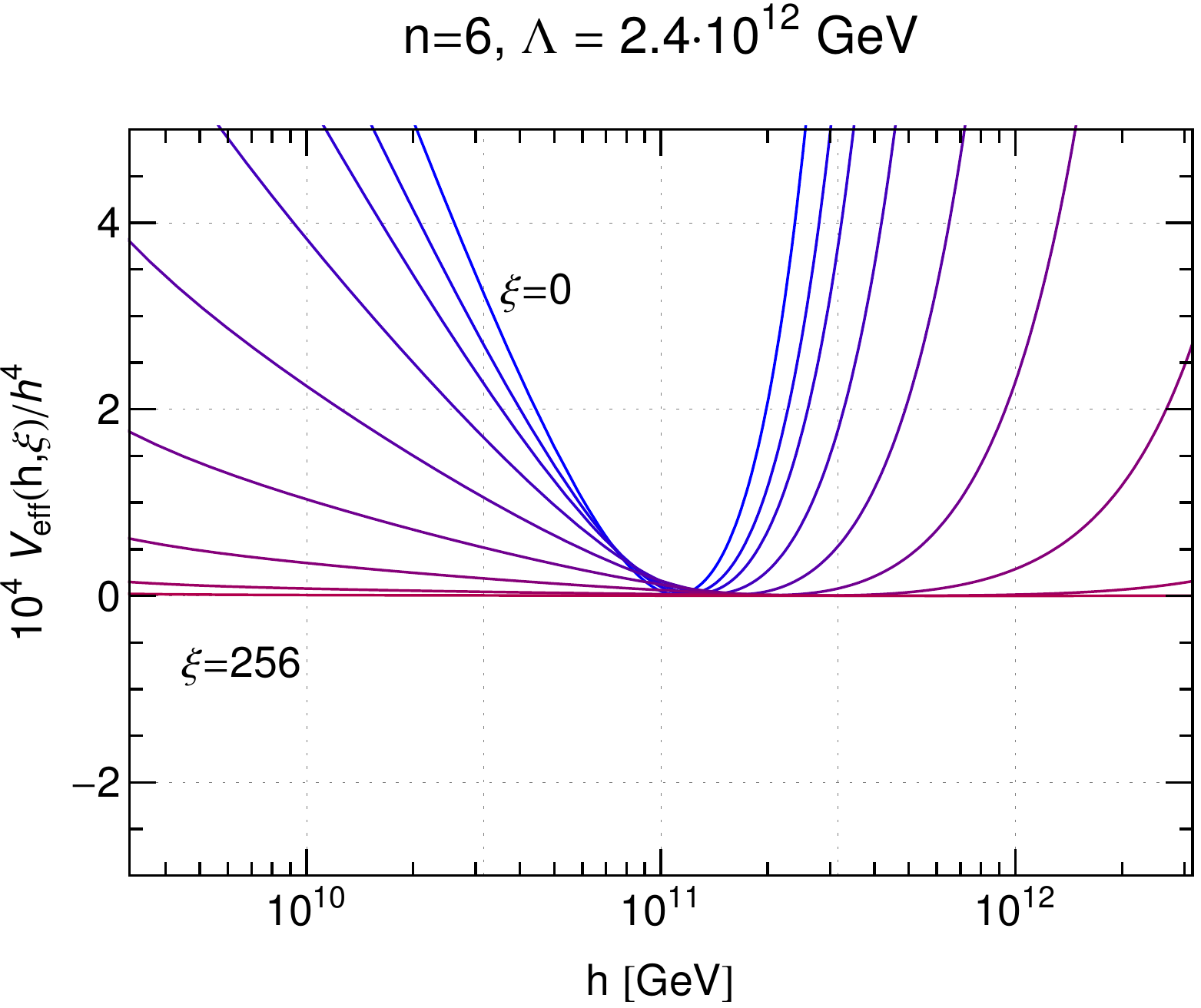}
  \includegraphics[width=0.45\textwidth]{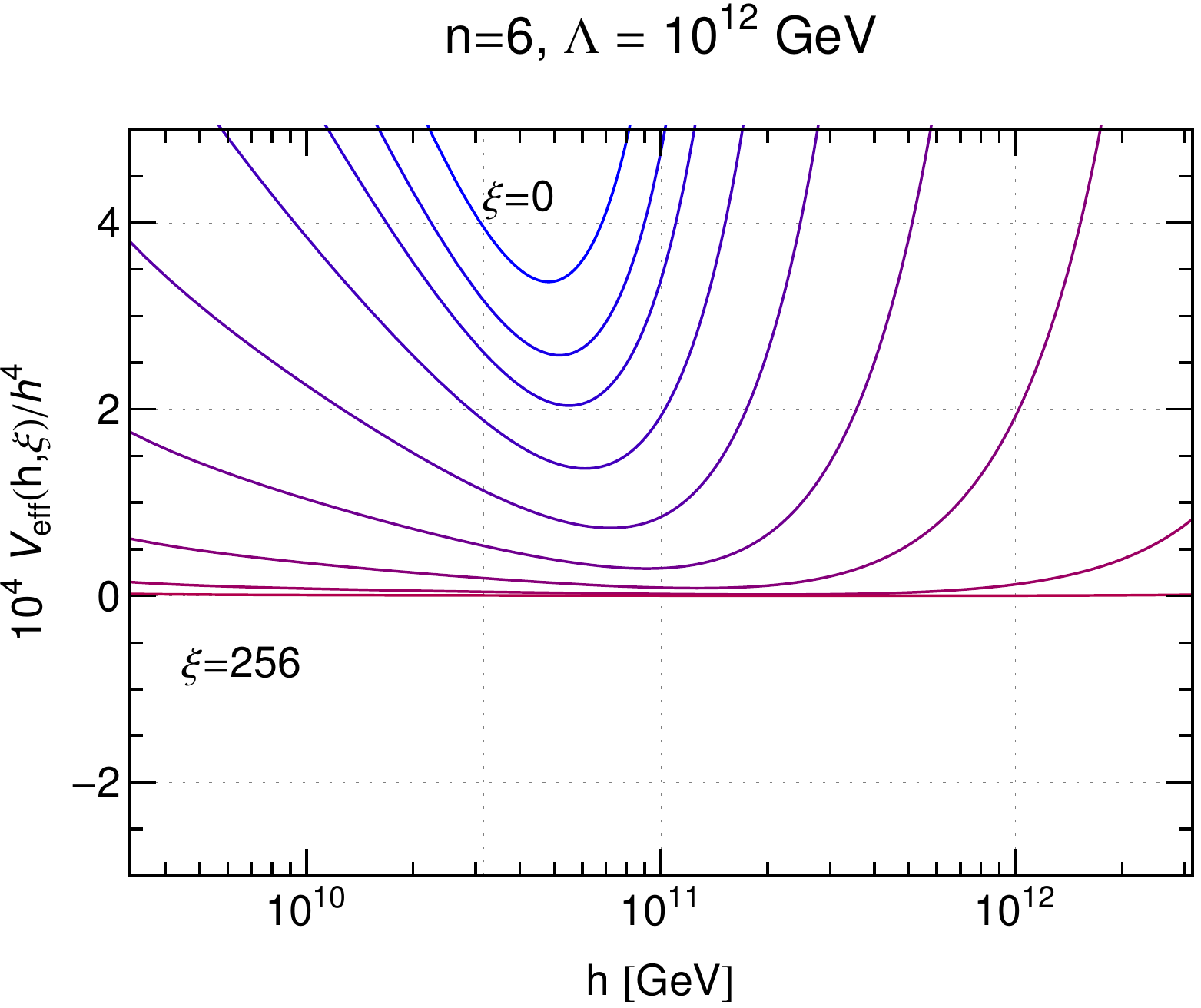}\\
\end{center}
\caption{\label{fig:Veff}%
\em
  Renormalization group improved (and IR resummed) effective potential $V(h,\xi)$ normalized to $h^4$ for the SM in Fermi gauge.
The different curves in each panel show the dependence on the gauge fixing parameter $\xi=0, 2, 4, 8, \dots, 256$. The four panels correspond to different choices of the scale $\Lambda$ of the dimension-six operator $\propto h^6/\Lambda^2$, as indicated.
}
\end{figure}

A close-up of the instability region of the RG-improved effective potential for $n=6$ is shown in Fig.\,\ref{fig:Veff}.  The upper left
plot corresponds to the Standard Model (without higher-dimensional operators, {\em i.e.}, $\Lambda\to\infty$) for several values of
$\xi$ from $0$ to 256. We see that the scale $h_I$ [at which $V(h_I)=0$] is indeed $\xi$-dependent. In the other plots we added to the potential the $c_6h^6/(8\Lambda^2)$ operator, with $c_6(\Lambda)=1$. For $\Lambda=5\times 10^{12}$\,GeV (upper right plot), the value of the potential at the high-scale minimum is negative, \emph{i.e.} deeper than the electroweak minimum. For $\Lambda=10^{12}$\,GeV the potential is positive at the high-scale minimum, such that the electroweak vacuum becomes the absolute minimum. For $\Lambda=\Lambda_{c}=2.4\times 10^{12}$\,GeV  both minima are exactly degenerate (lower left plot). It is apparent that this degeneracy occurs for all values of $\xi$ at the same value of $\Lambda_{c}$, as discussed above.\footnote{The value of the potential at the minimum is gauge invariant  for any value of $\Lambda$. The apparent gauge dependence in the upper and lower right plots is due to the fact that the minimum of the ratio $V/h^4$ shown in the plots is offset from the true minimum of $V$ for $V\not=0$.} On the other hand, the field value $h_c$ at which the minimum occurs does depend on $\xi$, as expected.

\begin{figure}[!t]
\begin{center}
  \includegraphics[width=0.8\textwidth]{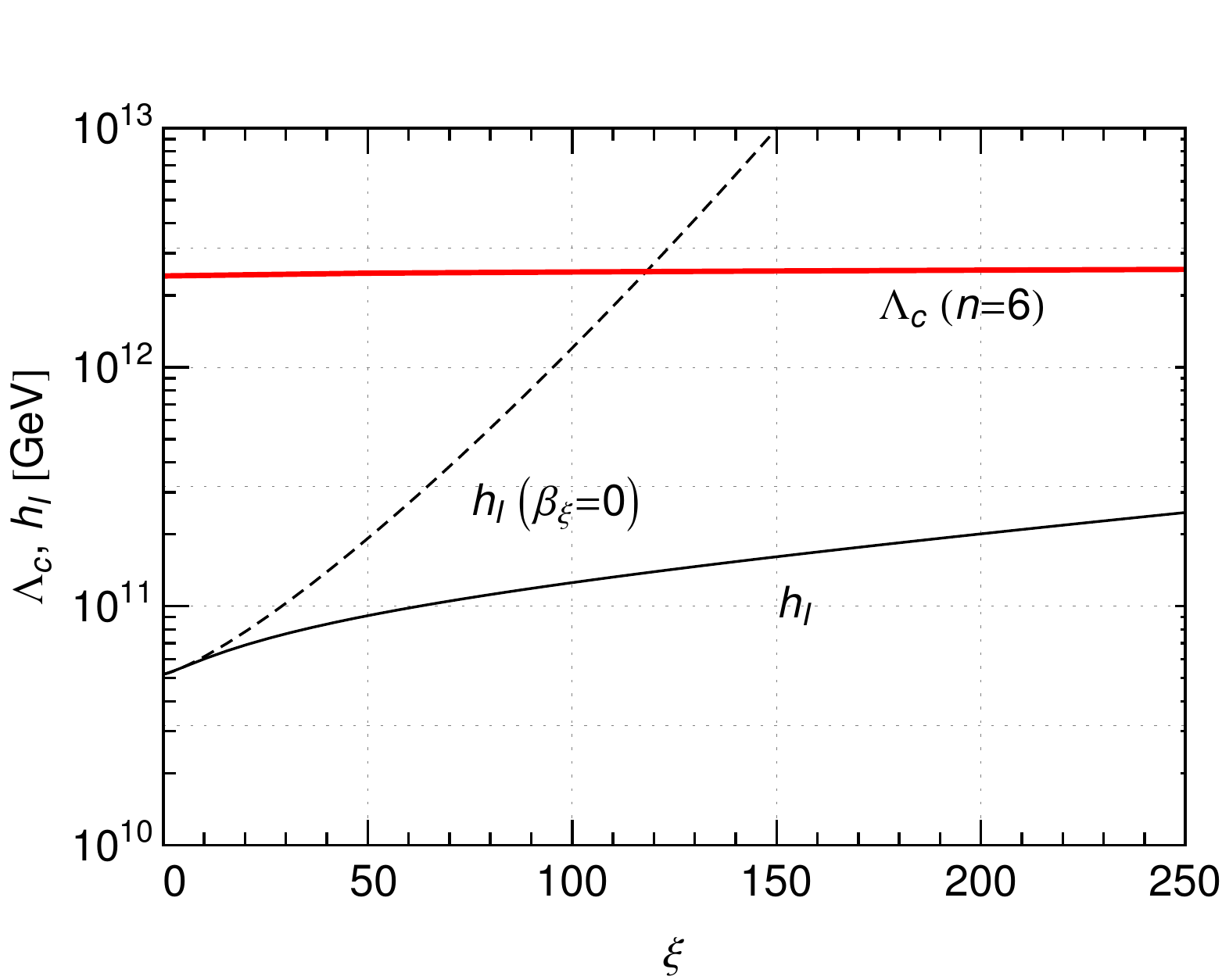}
\end{center}
\caption{\label{fig:instScale}%
\em
  Gauge (in)dependence of the instability scale of the Standard Model, as a function of the gauge fixing parameter $\xi$.
  The black lines reproduce the results from \cite{DiLuzio}, and correspond to the field value $h_I$ for
  which $V(h_I)=0$  (the dashed line is obtained if  the running of $\xi$  is neglected).  The red line shows the critical value $\Lambda_{c}$ of a dimension-six operator that stabilizes the SM potential giving a large scale minimum degenerate with the EW one. The residual $\xi$-dependence of $\Lambda_{c}$  is $\sim 6 \%$ within the range shown.
}
\end{figure}

In Fig.\,\ref{fig:instScale} we compare the critical scale $\Lambda_{c}$ obtained for $n=6$ with the instability scale $h_I$
at which $V=0$, comparing directly with the results in \cite{DiLuzio}, reproduced by the black lines.
Even when varying the gauge fixing parameter  over a large range,
$\xi=(0,250)$, we find that the $\Lambda_{c}$ scale determined from the RG improved one-loop potential is stable at the level of $6\%$, with the residual gauge dependence due to the perturbative  truncation of the potential. This is to be contrasted with the gauge dependence of the instability scale defined via the field value,
that varies by half an order of magnitude over the same range of $\xi$. Let us also mention that we checked that
{\em (i)} the effect of the running of the $c_{m>6}$ higher-dimensional operators generated by the running even when
only $c_6$ is present at the scale $\Lambda$, is minor, affecting $\Lambda_{c}$ by less than $2\%$. {\em (ii)} The value of $\Lambda_{c}$ is also stable at the percent level when varying (by a factor $2$ up or down) the renormalization scale
used to resum large logarithms.

\begin{figure}[!t]
\begin{center}
  \includegraphics[width=0.75\textwidth]{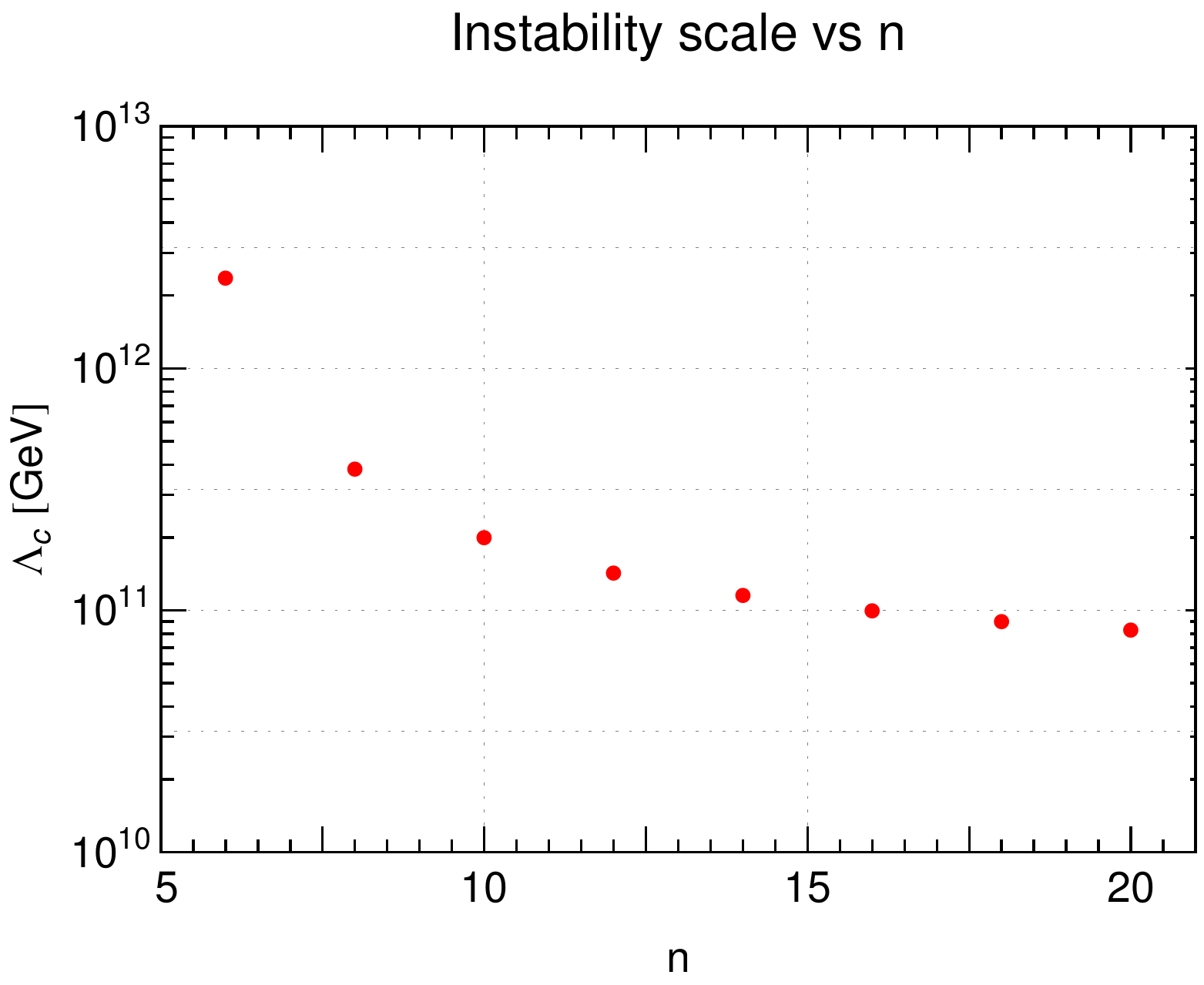}
\end{center}
\caption{\label{fig:instScaleVsDim}%
\em  Critical scale $\Lambda_{c}$ of a $h^n/\Lambda^{n-4}$ operator that stabilizes the SM potential giving a large scale minimum degenerate with the EW one, for various values of $n$.
The residual $\xi$-dependence for $\xi\in(0,250)$ is below the $10\%$ level for all $n$.
For large $n$ the critical scale approaches $\Lambda_I\simeq 10^{11}$\,GeV.}
\end{figure}

There is  freedom to choose the dimension $n$ of the non-renormalizable operator we use to extract the scale $\Lambda_c$. In Fig.\,\ref{fig:instScaleVsDim} we show the critical value $\Lambda_{c}$
obtained for various choices of $n=6,8,\dots$. For large $n$ the critical value asymptotes to $\Lambda_{I} \simeq 10^{11}$\,GeV. The reason is that
in this limit the effect of the higher-dimensional operator can be considered almost as a step function. Therefore, one may identify the asymptotic large-$n$ value, that we call  $\Lambda_{I}$ as the physical scale beyond which the electroweak vacuum becomes unstable. Irrespective of the precise physical interpretation of this scale, it is important
that it can be determined in an unambiguous way, independent of the gauge fixing. It is interesting to note that $\Lambda_{I}$
is actually rather close to the naive instability scale $h_I$ obtained in Landau gauge by demanding $V(h_I)=0$. As we show below this is not a coincidence but depends on some of the assumptions we have made, in particular for the scaling of $c_n$
with $n$.

Let us approximate the potential in the instability region as
\be
V(h, \Lambda)\simeq \frac14 \lambda(h) h^4 + c_n\frac{h^n}{2^{n/2}\Lambda^{n-4}}\ ,
\ee
where the bulk of the radiative corrections to the potential are captured in the running quartic $\lambda(h)$,  (that includes also finite one-loop contributions), evaluated at the renormalization scale $\mu=h$. This approximation is sufficient
to determine the leading scaling of $\Lambda_c$ and its limiting value $\Lambda_I$ with different parameters. In particular this approximation ignores the $\xi$-dependence of the radiatively corrected potential, but this is not an issue once we have a general argument for the $\xi$-independence of $\Lambda_c$. From the conditions $V(h_c, \Lambda_c)=0$ and $V'(h_c, \Lambda_c)=0$ we get 
\be
\lambda(h_c) \simeq\frac{\beta_\lambda(h_c)}{n-4}\, , \quad\quad
\Lambda_c \simeq h_c \left[\frac{-4nc_n}{2^{n/2}\beta_\lambda(h_c)}\right]^{1/(n-4)} \ .
\ee
Taking the large $n$ limit we see that $\lambda(h_c) \ra 0$, so 
that $h_c\to h_I$, the point at which the quartic coupling crosses zero: $\lambda(h_I)=0$. To take the large $n$ limit of $\Lambda_c$, which determines $\Lambda_I$, we need to specify how
$c_n$ scales with $n$. The generic expectation, based on simple $\hbar$ power counting, is that at tree-level $c_n\sim g^{n-2}$ where $g$ represents a generic coupling between the Higgs field
and the physics at the scale $\Lambda$. With this $n$ dependence
included we get the limit
\be
\Lambda_I\simeq\frac{g}{\sqrt{2}}h_I\ .
\ee
In the numerical analysis of Fig.~\ref{fig:instScaleVsDim} we simply
took $g=1$ but in general one does expect some dependence
on the coupling strength $g$. After all, in order for the new physics represented by the non-renormalizable operator to stabilize the potential, it should matter how strong is the coupling of that new physics to the Higgs field.\footnote{In fact, how the scale of  new physics able to stabilize the potential could vary depending on the coupling strength has been analyzed before, using particular models, see \cite{HungSher}.} Up to that unavoidable model-dependence we are nevertheless able to extract a gauge-independent scale to be very closely associated with the scale of instability of the SM potential.

In this context, it is interesting to ask  what is the highest scale where some (perturbative) new physics has to appear if one demands absolute vacuum stability. The highest scale is obtained for the lowest order $n=6$, and we take $c_6 \sim g^4$ in accordance with the expected scaling with the coupling strength to the new physics  [the validity of the effective description requires $g\gtrsim {\cal O}(1)$, so as to keep sufficiently separate $\Lambda$ and the potential minimum]. For the central value of the SM parameters one obtains (see Fig.~\ref{fig:Veff}) $\Lambda_c \lesssim 2.4\times 10^{12}$ GeV $\times\,\, g^2 \lesssim 3.8\times 10^{14}$ GeV, where we assumed $g<4\pi$ to get this rough estimate of the largest possible scale where new physics has to appear to ensure absolute vacuum stability.\footnote{The presence of the dimension-$5$ (Weinberg) operator with a similar suppression scale would not spoil this upper bound. As is well-known  \cite{stabnu}, heavy Majorana neutrinos tend to make the potential more unstable and therefore would call for even lower values of $\Lambda_c$.}

%%%%%%%%%%%%%%%%%%%%%%%%%%%%%%%%%%%%%%%%%%%%%%%%%%%%%%%%%%%%%%%%%%%%%%%%%%%%%
\section{Gauge Invariance and the Tunneling Critical Bubble\label{sec:Rc}}
%%%%%%%%%%%%%%%%%%%%%%%%%%%%%%%%%%%%%%%%%%%%%%%%%%%%%%%%%%%%%%%%%%%%%%%%%%%%%

In quantum field theory, false vacuum decay proceeds by the nucleation of 
bubbles that probe the lower-energy phase, are large enough to 
grow (with the bulk energy gain overcoming the surface tension), and eventually engulf the whole of space \cite{Voloshin}. The tunneling probability density (which determines the false vacuum lifetime) actually measures how likely it is to nucleate such bubbles per unit time and unit volume. 

This decay rate $\Gamma$ is calculated by first finding the so-called bounce solution $h_b$ to the Euclidean EoM \cite{Coleman}. This solution is $O(4)$ spherically symmetric, $h_b(\rho)$, with $\rho^2=t_E^2+r^2$, where $t_E$ is Euclidean time and $r^2=\vec{x}^2$. It has boundary conditions $h_b=v_f$
at $t_E\ra \pm\infty$ (where $v_f$ is the expectation value of the field in the false vacuum) and a turning point (assigned without loss of generality to $t_E=0$) with $\partial h_b/\partial t_E=0$ for all $\vec{x}$. The decay rate scales as $\Gamma\sim e^{-S_4}$, where $S_4$ is the Euclidean action for this bounce solution, $S_4=S_E[h_b]$  (see {\it e.g.} \cite{Rubakov} for a nice introduction). The gauge-independence of $S_4$ and $\Gamma$ has been discussed
in the literature before \cite{Metaxas,Plascencia}  and follows the usual pattern: the explicit $\xi$-dependence that appears in the effective action, described by the Nielsen identity (now for the Euclidean action):
\be
\xi\frac{\partial S_E}{\partial \xi} + \int d^4x K_E[h(x)] \frac{\delta S_E}{\delta h}=0\ ,
\label{NielsenE}
\ee
is compensated by an implicit dependence on $\xi$ of the solutions of the EoM/bounce equation, given by
\be
\xi\frac{d h_b}{d \xi} = K_E[h_b] \ ,
\label{dphibdxi}
\ee
and leading to 
\be
\frac{d S_4}{d \xi} = 0 \ .
\ee
Here,  rather than on $S_4$, we are interested in the radius of  the critical bubble, which provides another physical scale associated to the potential instability.
The critical bubble, the most likely bubble profile for vacuum decay, corresponds to the bounce solution evaluated at $t_E=0$, $h_B(r)\equiv h_b(\sqrt{0^2+r^2})$. It is an $O(3)$ spherically symmetric bubble configuration with zero total energy (as the tunneling process must conserve energy). In fact, the analytic continuation of the Euclidean bounce to Minkowski spacetime (with $\rho^2=t_E^2 + r^2\ra -t^2 + r^2$,
where $r^2=\vec{x}^2$ is the square of the distance to the bubble center) gives the real time evolution of the nucleated bubble (see \cite{CDL} for details).  
The field profile, $h_B(r)$, could be used to define the radius $R_c$ of the critical bubble, e.g. as the radius at which the field is half-way between the false vacuum value at infinity and its value at the center of the bubble, $h_B(R_c)=[h_B(0)-v_f]/2$. However, it is clear that such definition leads to a $\xi$-dependent $R_c$
as the bounce solution  depends on $\xi$ in a non-trivial way, according to Eq.~(\ref{dphibdxi}).\footnote{\label{foot:Rcalt} In practice, the $\xi$-dependence of the critical radius defined in this way might be small
in the SM case, given that $C(h)\propto h$ up to small logarithmic effects [see Eqs.~(\ref{CB0},\ref{CW0})]. However, we would like to give a definition of the critical radius that is $\xi$-independent from first principles.}

A $\xi$-independent critical radius can be defined if, instead of using the field bubble profile, one uses the energy density profile.
Let us show how this works using a derivative expansion approximation for the effective action, although it is clear that 
the general derivation does not rely on such expansion.

Let us assume then that the derivative expansion of the effective action converges and gives a sufficiently good approximation at second order in derivatives. The Euclidean effective action 
for the $O(4)$ symmetric bounce $h_b(\rho)$ reads
\be
S_E=2\pi^2\int_0^\infty\, d\rho\,\rho^3\left[\frac{1}{2}Z(h_b)(\partial_\rho h_b)^2+V( h_b)-V_f\right]\ ,
\ee
where $V_f=V( v_f)$ is the potential value in the false vacuum. The bounce solution $ h_b(\rho)$ satisfies the Euclidean EoM
\be
Z( h_b)\DAlambert_E  h_b+ \frac{1}{2}Z'( h_b)\  (\partial_\rho  h_b )^2 = V'( h_b)\ ,
\label{EoME}
\ee
with $\DAlambert_E  h_b\equiv \partial_\rho (\rho^3 \partial_\rho h_b)/\rho^3$
and depends on $\xi$ because both $Z$ and $V$ are 
gauge-dependent objects. Their gauge dependence is described by the corresponding Nielsen identities that are obtained by expanding the functional $K_E[ h(x)]$ of Eq.~(\ref{NielsenE}) in gradients\footnote{ To avoid confusion, note that our definition of $D(h)$ and $\tilde D(h)$ is different form that used in \cite{GK}.}
\be
K_E[ h(x)] = C( h) 
+ D( h) (\partial_\mu  h\ \partial^\mu  h)
+ \tilde D( h)  (\partial_\mu\partial^\mu  h) +O(\partial^4) \, ,
\ee
with spacetime indices contracted using the Euclidean metric.
As the Nielsen identity of Eq.~(\ref{NielsenE}) holds for
generic fields $ h(x)$, one can extract the $\xi$-dependence
of $V$ and $Z$ by  cancelling the terms of order $\partial^0$, $(\partial  h)^2 $ and $(\partial^2  h)$ separately. One gets the usual
relation for the effective potential
\be
\xi \frac{\partial V}{\partial \xi}+C( h,\xi) V'
=0 \, ,
\label{dVdxi}
\ee
where primes denote field derivatives, while for $Z$ one gets 
\be
\xi \frac{\partial Z}{\partial \xi} = - C  Z'
- 2 Z  C' -2DV'+2(\tilde D V')' \, .
\label{dZdxi}
\ee
That the functions $C$, $D$ and $\tilde D$ indeed fulfill these relations as a consequence of the Nielsen identity can be checked on a case by case basis (see e.g.~\cite{Metaxas,GK} and Appendix~C of \cite{EGK} where this is shown explicitly at one-loop for the Abelian Higgs model in Fermi gauge). 
In addition, the bounce solution depends on $\xi$ according to
Eq.~(\ref{dphibdxi}), which in this spherically symmetric case takes the form
\be
\xi\frac{d h_b(\rho)}{d\xi}= C( h_b) 
+ D( h_b) (\partial_\rho  h_b)^2
+ \tilde D( h_b) \DAlambert_E  h_b +O(\partial^4)\ .
\label{dphibdxiE}
\ee
Making use of the relations above it is then straightforward
to show that 
\be
\xi \frac{d }{d\xi}S_E[ h_b]=
0\ ,
\label{dSEdxi}
\ee
to order ${\cal O}(\partial^4)$, as expected.
%Both $V'$ and $V''$ can be shown to be of order $\partial^2$ by using the EoM in Eq.~(\ref{EoME}) or its $\rho$-derivatives.Therefore the terms in (\ref{dSEdxi}) are of order $\partial^4$ as indicated and $S_4=S_E[ h_b]$ is $\xi$-independent to the order we work. 
The proof
is extended to fourth order in the derivative expansion in 
Appendix~\ref{app:derexp} and of course holds in full generality
even when the derivative expansion is not applicable.

%On one hand side, these relations automatically ensure that physical observables, as for example the critical bubble size, will be gauge-independent. For example, a gauge-independent definition of the  critical bubble size (in leading order in gradients?) could be obtained by using the radius where the W-boson mass surpasses a certain value or where the scalar field $ h$ passes the barrier of the scalar potential. But notice also that the derivative expansion does typically not converge for the critical bubble.

As explained above, the critical bubble is given by $ h_B(r)=
 h_b(t_E=0,r)$, where $r$ is the distance in 3D space to the center of the bubble. The real time evolution of the bubble, after nucleation, is given by the analytic continuation of the $t_E=0$
bounce, $ h_B(t,r)= h_b(\sqrt{-t^2+r^2})$, which automatically satisfies the Minkowski EoM
\be
-Z( h_B)\DAlambert  h_B- \frac{1}{2}Z'( h_B)\  [(\partial_t h_B)^2-(\partial_r  h_B )^2] = V'( h_B)\ ,
\label{EoMM}
\ee
with $\DAlambert  h_B=\partial_t^2 h_B- \partial_r(r^2\partial_r h_B)/r^2$.  At $t=0$, $\partial_t h_B=0$ but $\partial_t^2 h_B=-\partial_r h_B/r$ so that Eq.~(\ref{EoMM}) does agree with the bounce equation (\ref{EoME}).
Notice also that the $\xi$-dependence
of $ h_B(r)$ follows directly from (\ref{dphibdxiE}) and is given by
\be
\xi\frac{d h_B(r)}{d\xi}= C( h_B) 
+ D( h_B) (\partial_r  h_B)^2
- \tilde D( h_B) \DAlambert  h_B+O(\partial^4)\ .
\label{dphiBdxi}
\ee
The energy density for the critical bubble is
\be
E_B(r) = \frac{1}{2}Z( h_B)(\partial_r  h_B)^2 +
V( h_B)-V_f\ .
\label{EBr}
\ee
Using Eqs.~(\ref{dVdxi}), (\ref{dZdxi}) and (\ref{dphiBdxi})
we get
\be
\xi\frac{d}{d\xi}E_B(r) = (\tilde D V')'(\partial_r h_B)^2
-\tilde D V' \DAlambert  h_B={\cal O}(\partial^4)\ ,
\ee
where, for the last step, we use the EoM for $ h_B$, Eq.~(\ref{EoMM}), and its $r$-derivative to show that $V'$ and $V''$ are quantities
of order ${\cal O}(\partial^2)$. This completes the check that the energy density profile is $\xi$-independent to the order we work.
The previous discussion, based on the derivative expansion at second order in derivatives, can be extended to higher orders as shown in Appendix~\ref{app:derexp}. 

Even when the derivative expansion is not convergent, the energy density profile must be $\xi$-independent in any self-consistent calculation: as shown in Section~\ref{sec:NI}, the energy-momentum tensor is $\xi$-independent, and therefore, the energy density of the critical bubble must be $\xi$-independent too. As the total energy density of the critical bubble integrates to zero,\footnote{\label{foot:E0}To prove this explicitly,  integrate by parts to show  $\int_0^\infty r^2 (V[h_B]-V_f)dr=-\int_0^\infty r^3 V' \partial_r h_B dr/3$ and use the EoM to substitute $V'$ and obtain in this way a relation between potential and surface energies.}
\be
E_{B,\text{tot}} = 4\pi^2\int_0^\infty r^2 E_B(r) dr =0 \ ,
\ee
with $E_B(r)$ starting negative at $r=0$, turning positive at some $r$ and going to zero at $r\ra\infty$, there must exist a distance $r=R_c$ at which $E_B(r)$ reaches a maximum. That is,
\be
\left.\frac{d E_B(r)}{dr}\right|_{R_c}=0\ ,\quad\quad
\left.\frac{d^2 E_B(r)}{dr^2}\right|_{R_c}<0\ .
\ee\label{rcritdef}
This value $R_c$ 
is our definition of radius of the critical bubble and is 
$\xi$-independent by construction. More explicitly, if we use the energy density expression in Eq.~(\ref{EBr}), taking its $r$-derivative and simplifying it using the field equation of motion, we arrive at the simple (implicit) result for the critical radius
\be
\label{Rcmax}
\left.R_c=\frac{3Z\partial_r h_B}{2 V'}\right|_{R_c}\ .
\ee
It can be checked, using the $\xi$-dependence of $Z,V$ and $h_B$
that indeed one has $dR_c/d\xi=0$, to  ${\cal O}(\partial^2)$.
After nucleation at $t=0$, the bubble radius can still be defined via the maximum of the energy density, which at $t>0$ includes also a nonzero contribution from the kinetic term $Z(h_B)(\partial_t h_B)^2/2$, and is $\xi$-invariant at all times. Although this definition is frame-dependent, for macroscopic values $t,r\gg R_c$, it can be shown that 
$R(t)\simeq \sqrt{R^2_c+t^2}\simeq t$ (in agreement with the fact that $h_B(r,t)$ is a function of $r^2-t^2$ \cite{CDL}).

\subsection{The Thin-Wall Critical Bubble} 
We discuss now the case of a thin-wall  critical bubble which, as usual, allows a good analytic understanding
of the parametric dependence of several quantities of interest.
The critical bubble wall is thin when the energy difference between
false and true vacua is small compared to the potential barrier that separates them. In this case the field profile of the bubble interpolates between the two vacua (true vacuum inside the bubble, false outside) with a very rapid transition, 
such that the wall thickness is much smaller than the bubble size.

Without making assumptions about the derivative expansion
we write the general expression for the Euclidean action for the bounce solution $ h_b(\rho)$ as 
\be
S_E=2\pi^2\int_0^\infty\, d\rho\,\rho^3\left[{\cal L}_K( h_b)+V( h_b)-V_f\right]\ ,
\ee
where ${\cal L}_K$ contains all the derivative terms beyond the simple $Z( h)(\partial h)^2/2$. We have to assume that the action $S_E$ is to a certain extent 'local'. For example the two-point function $\delta^2 S_E/\delta  h^2 $ falls off exponentially on a length scale of the order of the inverse Higgs mass. This does not imply that the derivative expansion converges since the bubble thickness is typically of the same scale.

In the thin-wall case it is easy to see heuristically that the critical radius should be $\xi$-independent: the effective action for the inner and outer parts of the bubble depend only on the constant pieces of the bounce profile. The $\xi$-dependence of these parts is given by the function $C$ as in Eq.~(\ref{dphibdxiE}) evaluated for a constant field value, with negligible contributions from the derivative terms (important only in the bubble wall). Since $\xi$-changes map the inner and outer parts of the critical bubble onto themselves, the bubble size is then  $\xi$-independent up to the negligible  bubble wall thickness.

This is of course consistent with the analytic estimate of the critical radius in the thin-wall case. The total Euclidean action receives a contribution from the inside region of the bubble which is approximately given by
\be
S_{E,{\rm in}} \simeq 2\pi^2\int_0^{R_c} d\rho\,\rho^3\left[V( h_b)-V_f\right]\ \simeq -\frac12 \pi^2 R_c^4  \, \epsilon_V \, , 
\ee
where $\epsilon_V\equiv V_f-V_t$, is the potential difference between the true and false minima. On the other hand, the wall
contributes an amount proportional to the surface of the critical bubble. We can write it in terms of the wall tension
as 
\be
S_{E,{\rm wall}} = 2\pi^2\int_{R_c-\Delta}^{R_c+\Delta} d\rho\,\rho^3\left[{\cal L}_K( h_b)+V( h_b)-V_f\right]\equiv 2 \pi^2 R_c^3  \, \sigma \, ,
\ee
which can be considered as the definition of $\sigma$, and gives the right scaling with $R_c$. All the (possibly complicated) dependence of ${\cal L}_K( h_b)$ on field gradients is modelled
by the wall tension. Finally, the exterior of the bubble, for which $V(h_b)-V_f\simeq 0$ gives a zero contribution to the total action,
$S_{E,{\rm out}}\simeq 0$.

Extremizing the Euclidean action $S_4=S_{E,{\rm in}}+S_{E,{\rm wall}}+S_{E,{\rm out}}$ with respect to the size of the critical bubble then yields the standard results
\be
R_c=\frac{3\sigma}{\epsilon_V}\ ,\quad
S_4 = \frac{27 \pi^2 \sigma^4}{2\epsilon_V^3}\ .
\label{Rc}
\ee
Obviously $\epsilon_V$ is gauge-independent, as the values of the potential at its minima are $\xi$-independent. Since the action $S_4$ is also $\xi$-independent, this is then also true for 
the wall tension $\sigma$ and the size of the critical bubble $R_c$.
The same result for $R_c$ can be obtained by examining the total energy of the critical bubble
\be
E_{B,\text{tot}} = 4\pi\int_0^\infty r^2 E_B(r) dr \simeq
4\pi\left(\sigma R_c^2-\frac13 \epsilon_V R_c^3\right)=0\ ,
\ee
which can be solved for $R_c$ with the same result as in (\ref{Rc}).
This critical radius coincides with the location of the maximum in the energy density profile (that we used to define $R_c$ in the general case)
as 
\be
E_B(r)\simeq \left\{\begin{array}{ccl}
-\epsilon_V && (r<R_c)\ ,\\
\sigma\delta(r-R_c) && (r=R_c)\ ,\\
0 && (r>R_c)\ ,
\end{array}\right.
\ee
peaks at $r=R_c$.

\subsection{The SM Case}

The energy scale $\mu_d\sim 1/R_c$ associated to vacuum decay and determined by the critical bubble radius is of relevance for discussing  the impact of heavy physics beyond the SM on the lifetime of the EW vacuum. In particular, if new physics starts to be relevant at such scale, then it can impact the lifetime of the false vacuum even if $\mu_d$ were much higher than the instability scale $\Lambda_I$. In fact this is precisely what happens
in the SM.

As we have seen, at field values much higher than the EW scale the SM effective potential can be well approximated as
\be
V(h)\simeq \frac14 \lambda_{\rm eff} h^4\ ,
\ee
with $\lambda_{\rm eff}<0$. As is well known \cite{LW}, the bounce solution
for such a potential can be calculated analytically (assuming a constant $\lambda_{\rm eff}$) and is
\be
h_b(\rho)=\frac{2\sqrt{2}}{R\sqrt{-\lambda_{\rm eff}}}\left(\frac{1}{1+\rho^2/R^2}\right)\ ,
\label{quarticbounce}
\ee
where $R$, that determines the critical bubble size, is arbitrary.
For any value of $R$ one gets the same Euclidean action for the bounces, with $S_E[ h_b]=8\pi^2/(-3\lambda_{\rm eff})$.
The energy profile for the critical bubble is
\be
E_B(r) = \frac{16(r^2/R^2-1)}{(-\lambda_{\rm eff})R^4(1+r^2/R^2)^4}\ ,
\ee
and has a maximum at $r=\sqrt{5/3}R\simeq 1.3 R$ so that $R$
basically coincides with the critical radius defined through the maximum of $E_B(r)$.

As appreciated long ago \cite{Arnold}, the scale invariance of the bounce solution is broken by radiative effects: $\lambda_{\rm eff}$ is not really constant but receives potentially large radiative corrections enhanced by powers of large logarithms $\propto \log(h/\mu)$, where $\mu$ is the renormalization scale. In order to resum such 
logarithms one chooses $\mu\simeq h$, in practice evaluating a
running $\lambda_{\rm eff}(\mu)$ at $\mu=h$. The 
Euclidean action that suppresses the EW vacuum decay is minimal
at the field value (or renormalization scale) at which 
$\lambda_{\rm eff}$ reaches its most negative value, that is,
$d\lambda_{\rm eff}/d\log\mu\equiv \beta_{\lambda_{\rm eff}}=0$, with $\lambda_{\rm eff}<0$. The scale at which that happens
singles out one special size for the critical bubble, with $1/R_c\sim 
\mu_d$ and $\beta_{\lambda_{\rm eff}}(\mu_d)=0$.

\begin{figure}[t]
\begin{center}
  \includegraphics[width=0.75\textwidth]{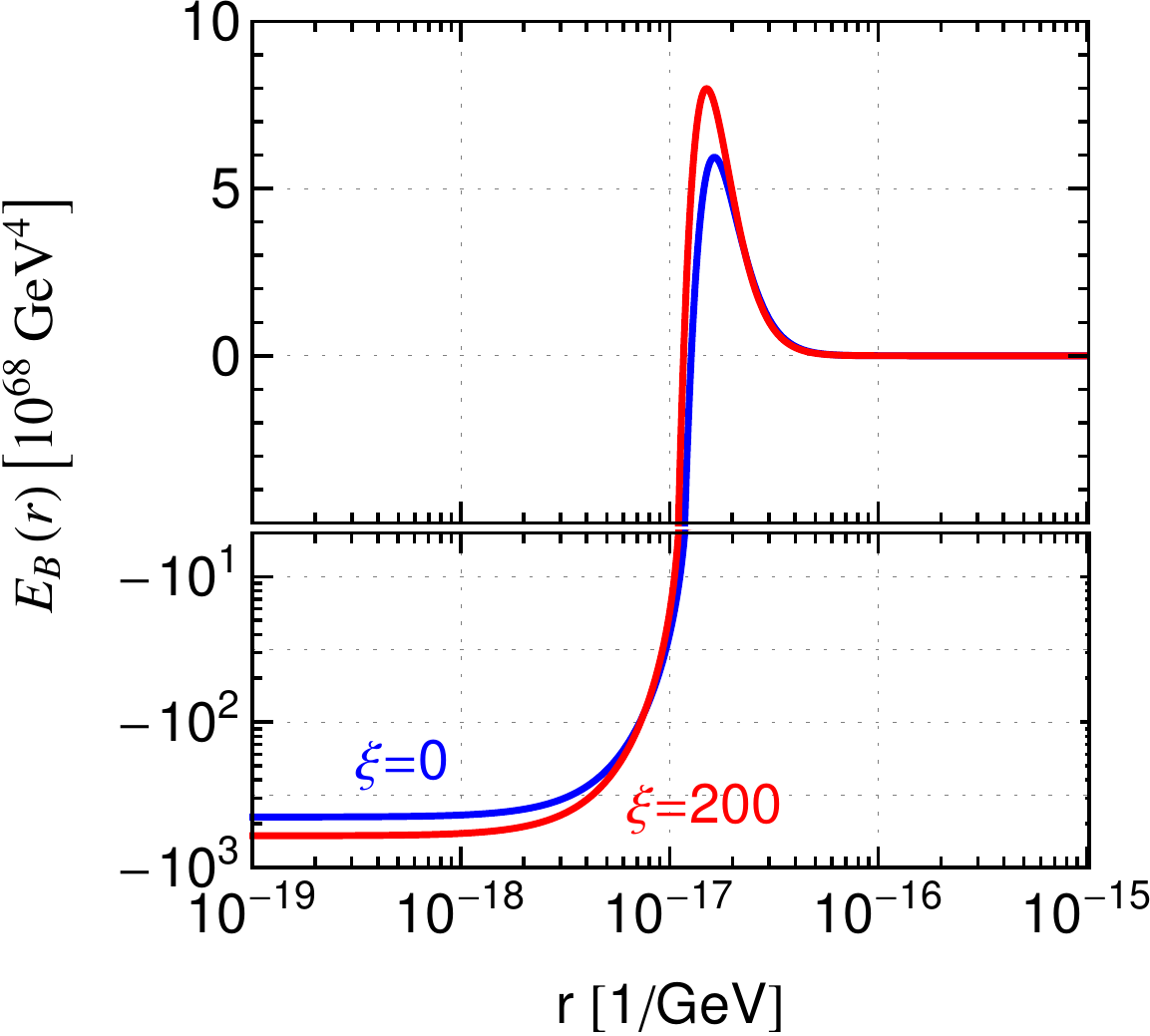}
\end{center}
\caption{\label{fig:profile}%
\em Spatial profile of the energy density (\ref{EBr}) for the critical bubble for two different values $\xi=0, 200$ in Fermi gauge.  These results are obtained solving the bounce equation (\ref{EoMM}) numerically (using the RG-improved and IR resummed 1-loop effective potential and the corresponding correction to the kinetic term) and choosing the field value in the bubble center $h_B(r=0)$ such that the bounce action is minimized.
}
\end{figure}

\begin{figure}[th]
\begin{center}
  \includegraphics[width=0.75\textwidth]{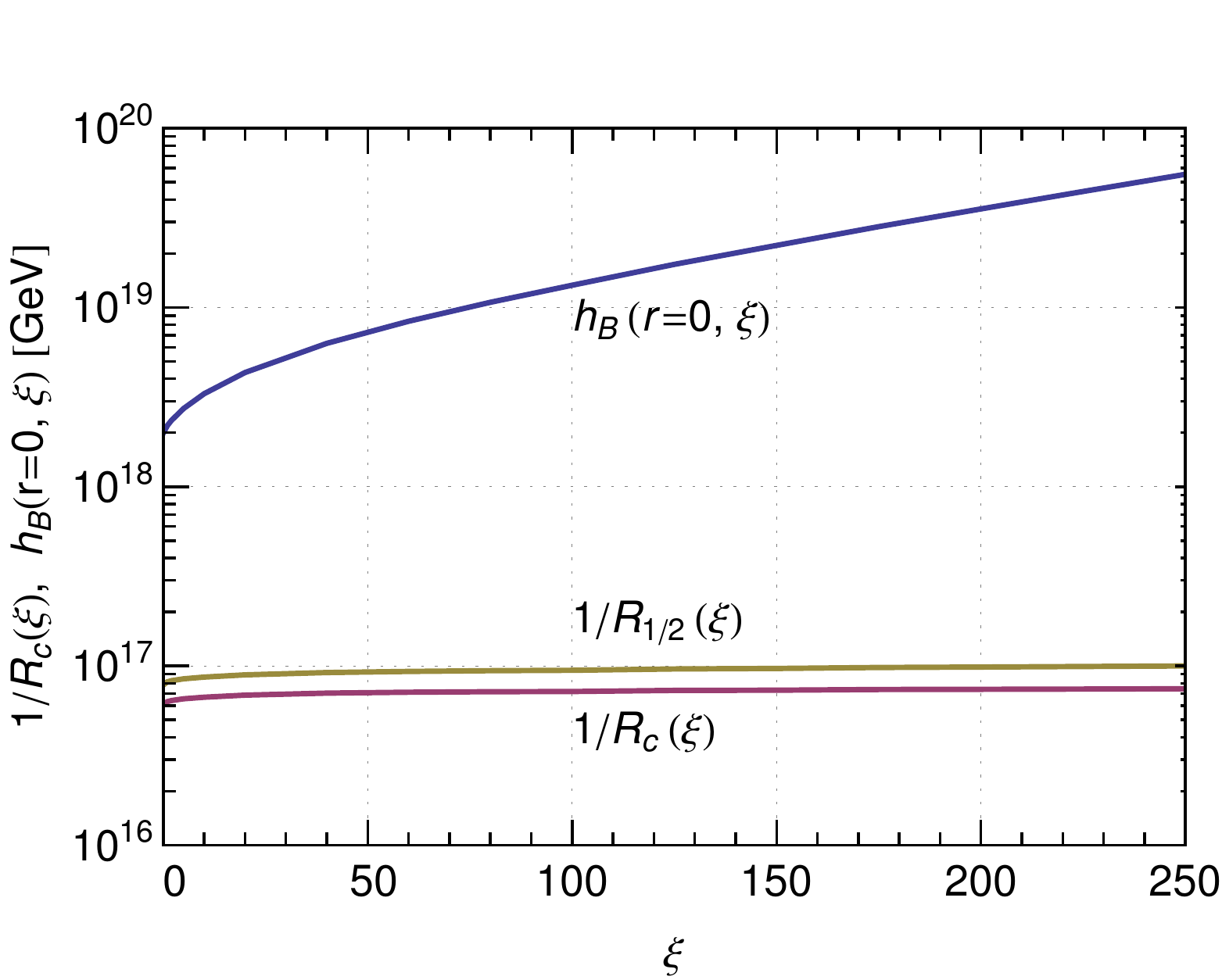}
\end{center}
\caption{\label{fig:rCrit}%
\em Dependence of the inverse critical radius $1/R_c$ and the field value at the center of the critical bubble $h_B(r=0)$ on the 
gauge-fixing parameter. 
The bubble radius varies at the $10\%$ level for  $\xi=0, 200$, consistent with the expected gauge independence and residual perturbative uncertainties. The value $h_B(r=0)$, on the other hand, is gauge dependent.
For comparison we also show the radius $1/R_{1/2}$ at which the field value has dropped to $h_B(r=0)/2$, which in this particular case is nearly $\xi$-independent also, see footnote~\ref{foot:Rcalt}.
}
\end{figure}
For the measured $M_h$ and $M_t$ (at the central values of their experimental intervals) one gets \cite{stab0} $\mu_d\simeq 10^{17}$ GeV, to be compared with the instability scale $\Lambda_I\simeq 10^{11}$ GeV. 
 Figure~\ref{fig:profile} shows the energy-density profile of the critical bubble, for two values of $\xi$ illustrating also the $\xi$-independence of the profile. The critical radius derived from it, is therefore also $\xi$-independent, as show in Figure~\ref{fig:rCrit}, as a function of $\xi$. For comparison, notice that the value of the field at the center of the bubble is $\xi$-dependent, as expected.
The hierarchical separation between $\mu_d$ and $\Lambda_I$  is a consequence of the nearly scale independence of
the effective potential and the Euclidean action for the bounce (see \cite{Gino,EspinosaTOP2015} for some recent discussions on this).
From this fact it follows that BSM physics modifying the effective potential at the scale $\mu_{BSM}$ with $\mu_{BSM}\leq \mu_d$ can have a large impact on the EW vacuum lifetime even if
$\mu_{BSM}\gg \Lambda_I$. 

The simplest (and best motivated) example of such scenario has been discussed long ago: a see-saw scenario with heavy right-handed neutrinos of mass $M_{\nu_R}$ \cite{stabnu}. This model requires large neutrino Yukawa couplings (to fit light neutrino masses $m_\nu$) that cause the running 
$\lambda_{\rm eff}$ to become even more negative above the $M_{\nu_R}$ threshold, further destabilizing the EW vacuum. In fact, the prediction of a too-short lived EW vacuum can be used \cite{stab2.5} to rule out
$M_{\nu_R}\geq 10^{13-14}$ GeV  for $m_\nu\simeq 0.1-1$ eV. 

In principle, even new physics at $\mu_{BSM}> \mu_d$ could impact the EW vacuum lifetime, as recently emphasized in \cite{Branchina}. All that is required is that the BSM physics makes the Higgs effective potential even more unstable at $h>\mu_d$
(and this implies already Planckian physics) so that bounces with even smaller Euclidean action are possible. However, it is difficult to find plausible reasons why physics at the Planck scale (presumably more fundamental) should worsen the potential instability (unlike what happened in the well-motivated see-saw scenario where one understands the origin of the instability as due to new Yukawa couplings). In connection to this, it has also been noticed \cite{EspinosaTOP2015} that, while BSM physics with $\mu_{BSM}\gg \Lambda_I$ can easily make the EW vacuum more unstable, it is much harder to make the potential more stable. The reason is that increasing the Euclidean action at high scales suppresses bounces that cause tunneling to such large scales but does not modify 
the bounce at the scale $\mu_d$, which still dominates vacuum decay and therefore leads to the standard EW vacuum lifetime.

\subsection{Including Gravity}
 
We now extend our considerations by including gravity \cite{CDL,gravdecay2}. This is particularly relevant since in the SM the tunneling bubbles do not experience a flat geometry, but an  anti-de Sitter one. For discussions on the effect of gravitational corrections on the EW vacuum decay see \cite{IRST,EFT,MPW,Branchinagrav,Rajantie,SSTU}. We start by writing the most general rotationally
invariant Euclidean metric
 \be
 ds^2=d\zeta^2+\rho^2(\zeta)d\Omega_3^2,
 \ee
 where $\zeta$ is a radial coordinate measuring distances along radial curves normal to three-spheres, $d\Omega_3^2$ is the line element angular distance on a unit $3$-sphere and $\rho$ is the radius of curvature of each three-sphere. Of this metric we need the corresponding Ricci scalar
 \be
 \label{a}
 R=\frac{6}{\rho^2}\left(-\rho\rho''-\rho'^2+1\right),
 \ee
 where primes stand for differentiation with respect to $\zeta$. 
 The EoM/bounce equation for $h$ is $\xi$-independent also in curved space-time. The reason is that the background metric is obtained through Einstein's equations and is sourced by the energy-momemtum tensor. Being the latter  $\xi$-independent, so is the metric. This implies that the EoM in curved space-time differs from the flat space-time one only by an object, the metric, that is invariant under $\xi$ changes.
 
 Vacuum decay is controlled by the difference $S_{E}[h_b]-S_{E,f}$ between two Euclidean actions, the one for the bounce and the background one in the false vacuum. 
 The bounce action can be written as
 \be
 \label{aa}
 S_E[h_b]=2\pi^2\int d\zeta\,\rho^3\left({\cal L}_K+V-\frac12 m_P^2 R\right),
 \ee
 where $m_P^2=1/(8\pi G)$ is the reduced Planck mass (squared) and $G$ is Newton's constant. ${\cal L}_K$ and $V$ are the Lagrangian containing the derivative terms and the effective potential term, respectively. Although our results below hold beyond the derivative expansion, for simplicity we will take ${\cal L}_K=Z(h){h'}^2/2$, with $h'\equiv dh/d\zeta$ and give below many expressions simplified to this order of the expansion. The background action is
 $S_{E,f}=-24\pi^2 m_P^4/V_f$ for a de Sitter false vacuum ($V_f>0$, where $V_f$ is the false vacuum potential) as in the SM case and $S_{E,f}=0$ for a Minkowski one.
 
By inserting Eq. (\ref{a}) into Eq. (\ref{aa}), we obtain
 \begin{eqnarray}
 \label{aaa}
 S_E[h_b]&=&2\pi^2\int d\zeta\,\left[\rho^3\left({\cal L}_K+V\right)+3m_P^2\left(\rho^2\rho''+\rho \rho'^2-\rho\right)\right]\nonumber\\
 &=&2\pi^2\int d\zeta\,\left[\rho^3\left({\cal L}_K+V\right)-3m_P^2\rho\left(1+ \rho'^2\right)\right],
 \end{eqnarray}
 where in the last step we have simplified the curvature term by integrating by parts.\footnote{Possible boundary terms play no role in this calculation and we simply ignore them throughout.}
 From this action one gets the Euclidean EoM for the field, which gives the bounce equation
\be
Z h''+ 3\frac{\rho'}{\rho} Z h' +\frac12 Z' {h'}^2 = V' \ ,
\label{bounceCDL}
\ee
where, with an abuse of notation we use $Z'\equiv \partial Z/\partial h$
and $V'\equiv \partial V/\partial h$, while $\rho'\equiv d\rho/d\zeta$ and $h'\equiv dh/d\zeta$.

The next step is to employ the $\zeta\zeta$-component of the Euclidean Einstein's equations
 \be
 \label{bb}
 G_{\zeta\zeta}=\frac{3}{\rho^2}\left(1-\rho'^2\right)=
  -\frac{1}{m_P^2}\,T_{\zeta\zeta}\ ,
 \ee
 where $G_{\zeta\zeta}$ is the  $\zeta\zeta$-component of Einstein's tensor and $T_{\zeta\zeta}$ is the  $\zeta\zeta$-component of the energy-momentum tensor,  with $T_{\zeta\zeta}=Z {h'}^2/2-V$.  
Using now Eq. (\ref{bb}) in Eq. (\ref{aa}), we can further simplify the action as
\bea
 S_E[h_b]&=&2\pi^2\int d\zeta\, \left[
\rho^3\left({\cal L}_K+V- \,T_{\zeta\zeta}
\right) -6\rho\, m_P^2 \right]\nonumber\\
&=& 4\pi^2\int d\zeta\, \left(
\rho^3V -3\rho\, m_P^2 \right)\ .
 \eea

The important remark at this point is that, since the  energy-momentum tensor is 
$\xi$-independent,  so must be Einstein's equations and the various components of the metric.   
Having previously shown that the effective action calculated on the bounce is $\xi$-independent provided that it is evaluated on configurations such that $\xi \,d h_b/d\xi=K[h_b]$, by reasoning as in the flat case, we conclude that the action is $\xi$-independent. The background action, $S_{E,f}=-24\pi^2 m_P^4/V_f$, is trivially $\xi$-independent as the potential value in the false vacuum, $V_f$, is $\xi$-independent.  

 Concerning the $\xi$-independence of the critical bubble radius in the presence of gravity, one can still use a bubble profile that is $\xi$-independent if, instead of using the field itself, one resorts to the profile of $T_{\zeta\zeta}$. Although defining a local energy density is in general not possible in the presence of gravity,
spherical symmetry allows one to give a well-behaved definition \cite{GRAVITATION}.
The total energy/mass of the critical bubble is given by the integral
(see {\it e.g.} \cite{MPW} for a recent discussion)
\be
\label{Egrav}
E = 4\pi \int_0^\infty d\rho\, \rho^2\, T_{\zeta\zeta}=
 4\pi \int_0^\infty d\rho\, \rho^2\, \left(\frac12 Z {h'}^2 + V-V_f\right)
\ ,
\ee
where $T_{\zeta\zeta}$ is now the Minkowski one.
Although $T_{\zeta\zeta}$ comes from matter alone, $E$ includes
gravitational self-interaction contributions (a similar formula is used in defining the total mass of a spherically symmetric star). 
In this respect, it is instructive to apply (\ref{Egrav}) to the thin-wall case taking the radius $R$ of the bubble as arbitrary (and setting $Z=1$ and $V_f=0$ here for simplicity). Defining the wall tension
as
\be
\sigma\equiv \int_0^\infty d\xi\left(\frac12 {h'}^2 + V\right)\ ,
\ee
we can use ${h'}^2/2 + V=\sigma \delta(\xi-\bar\xi)$, where
$\bar\xi$ corresponds to the position of the bubble wall, $R=\rho(\bar\xi)$. The total mass/energy of
such bubble is then
\bea
E(R) & =& 4\pi \int_0^R d\rho\,\rho^2 (-\epsilon_V) +4\pi\int_{\bar\xi-\delta}^{\bar\xi+\delta} d\xi\, \rho^2 \rho'\,\sigma \delta(\xi-\bar\xi)\nonumber\\
&=&-\frac{4\pi}{3}\epsilon_V R^3 + 4\pi\sigma R^2 \frac12(\rho'_++\rho'_-)\ ,
\eea
where $-\epsilon_V$ is the value of the potential inside the bubble, and, in evaluating the second (wall) integral we have taken care of the fact that $\rho'$, given by Eq.~(\ref{bb}), jumps at the wall with
\bea
\rho'_-&\equiv &\rho'(\bar\xi-\delta) = \sqrt{1+8\pi G \epsilon_V R^2/3}\nonumber\\
\rho'_+&\equiv &\rho'(\bar\xi+\delta) = \sqrt{1-2GE(R)/R}\ ,
\eea
with $\rho'_+$ corresponding to the Schwarzschild solution with total mass $E(R)$. Solving for $E(R)$ one gets
\be
E(R) = -\frac{4\pi}{3}\epsilon_V R^3 +4\pi \sigma R^2\sqrt{1+8\pi G\epsilon_V R^2/3}-8\pi^2 G\sigma^2 R^3\, 
\ee
which shows explicitly the gravitational contributions to the total energy of the bubble (and reproduces the result given in \cite{SW}). 

The critical radius in this thin-wall case can then be obtained by solving $E(R_c)=0$, with the result
\be
R_c=\frac{R_{c,0}}{1-2\pi G \epsilon_V R_{c,0}^2/3}\ ,
\ee
where $R_{c,0}=3\sigma/\epsilon_V$ is the critical bubble radius without gravitational effects, as in Eq.~(\ref{Rc}). This $R_c$ agrees with the original result of Coleman-De Luccia \cite{CDL} and shows how gravitational effects force $R_c>R_{c,0}$.
This result highlights one important difference with respect to the flat case: gravity contributes a  negative energy density inside the bubble, where the  geometry of space is distorted and for large bubble radius the enclosed volume grows only like $R^2$ rather than $R^3$. While energy conservation requires  critical  bubbles to always have zero energy, it is not guaranteed that a critical bubble exists (just making its radius large enough) so that it becomes possible that  the gravitational contribution prevents the tunneling to happen through critical bubbles \cite{CDL}.

It is straightforward to show that the total energy of the critical bubble \cite{MPW} in the general case, without the thin-wall assumption, as given in Eq.~(\ref{Egrav}), is zero. Simply integrate by parts as explained in footnote~\ref{foot:E0}, paying attention now to the $\zeta$-dependence of $\rho$ and use the bounce equation (\ref{bounceCDL}) to arrive at a relation between the kinetic and potential contributions, that ensures $E=0$. The radius defined as the maximum of $T_{\zeta\zeta}$, $\partial_\rho T_{\zeta\zeta}=
(1/\rho') \partial_\xi T_{\zeta\zeta}
=0$ is then obtained, after using the bounce equation (\ref{bounceCDL}), as
\be
R_c=\left.\rho'\frac{3Zh'_B}{2 V'}\right|_{R_c}\ ,
\ee
which generalizes the result of Eq.~(\ref{Rcmax}) including gravity effects. Here
\be
\label{Rcgen}
\left.\rho'\right|_{R_c}=\sqrt{\left.1+\frac{8\pi G}{3} R_c^2\left(\frac12 Z{h_B'}^2-V\right)\right|_{R_c}}
\ .
\ee
We see once again that $R_c$ thus defined is indeed $\xi$-independent, as can be explicitly checked using the $\xi$-derivatives of $Z$, $V$ and $h_B$ extracted from the Nielsen identity.\footnote{In the derivation above we have implicitly assumed that $\rho'\neq 0$ , which is correct when the false vacuum is Minkowski. In the case of a dS false vacuum (like the SM EW vacuum) one can have $\rho'=0$ at some $\zeta$ with important implications for the bounce properties and the vacuum decay rate. For a thorough discussion of this point see \cite{EFT,MPW}.} 

Concerning the impact of gravity on $R_c$ it would be wrong to 
identify the $\rho'$ factor as the only effect, as the value of the ratio $3Zh'_B/(2V')$ has to be evaluated at $R_c$, not $R_{c,0}$, and also because gravity affects the bounce equation (\ref{bounceCDL}), thus modifying $3Zh'_B/(2 V')$ as a function of 
radial distance.

Finally, when gravitational effects make a false dS vacuum stable and there is no possibility of vacuum decay via nucleation of bubbles of the true phase there is still the possibility of decay via Hawking-Moss instantons \cite{HM}. In that case the vacuum decay is controlled by the value of the potential at the maximum separating the two vacua, and this being a $\xi$-independent quantity so is the Hawking-Moss decay rate. There is no radius associated to this type of instanton, or more precisely, the transition occurs in a whole
Hubble patch, of size determined by the false vacuum value, $V_f>0$, again a $\xi$-independent quantity.

\section{Stabilization by Thermal Effects\label{sec:Tc}}

As is well known, thermal corrections tend to restore broken gauge symmetries \cite{Tsym,DJ}. It is therefore natural to ask at what
temperature the deep minimum of the potential at high field values
would be made degenerate with the low-scale minimum (that at such high $T$ would already be at the origin $h=0$ in field space) by thermal corrections.\footnote{Usually the thermal change of the potential, e.g. across a phase transition, is pictured keeping the origin fixed while the broken minimum is raised as $T$ increases. A more correct depiction should keep the potential at $h\gg T$ fixed (as thermal corrections in that field range are Boltzmann suppressed) while the symmetric minimum gets deeper as $T$ increases. Of course both pictures are directly related by a field-independent term in the potential, but that term is $T$-dependent and physical.} That critical temperature $T_c$ provides yet another physical scale associated with the potential instability.

The proof that $T_c$ is $\xi$-independent requires the generalization of the Nielsen identity for the thermally corrected effective potential. The fact that Nielsen identities also hold at finite temperature (suitably modified to take thermal effects into account)
ultimately follows from the fact that the partition function respects
the BRST symmetry so that  Ward identities still hold at finite $T$. The same applies to the Nielsen identity, that can be regarded as a Ward identity  for the effective potential. It is in fact straightforward to generalize the Nielsen identity (\ref{NIV}) to include thermal corrections (see e.g. \cite{Das,GK}). One has
\be
\xi \frac{\partial V_T}{\partial \xi}+C_T( h,T,\xi) V_T'
=0 \, ,
\label{dVdxiT}
\ee
where $V_T=V( h,T,\xi)$ is the thermally corrected potential
and $C_T( h,T,\xi)$ generalizes to finite temperature the function $C( h,\xi)$ of Eq.~(\ref{NIV}). Appendix~\ref{app:NIT} gives explicit details about the calculation of $V_T$ and $C_T$ in the SM.

Armed with the Nielsen identity, we can again interpret it as
telling how the explicit $\xi$-dependence of the potential $V_T$
is compensated by an implicit $\xi$-dependence of the field given
by
\be
\xi\frac{d h}{d\xi} = C_T( h,T,\xi)\ ,
\ee
so that changing $\xi$ is equivalent to a field redefinition that does not change the physics. It is then clear that the value of the critical temperature, determined by the degeneracy of the potential at its
two minima, is $\xi$-independent: a change of $\xi$ modifies the location of the minima but not the value of the potential there.

In the case of the critical $T$ for the SM instability, its numerical value turns out to be many orders of magnitude larger than the instability scale for the central value of $M_h$ and $M_t$, that we take to be a typical case. We find $T_c\simeq 10^{29}$ GeV, which is even higher than the Planck scale and therefore  of little interest:
at $m_P$ one certainly expects gravitational physics to change the potential in any case. The fact that $T_c$ is so large is related to the fact that
we are probing the potential at the scale associated to the non-standard minimum, and this scale is much larger than the instability
scale (in fact the scale of the minimum is of order $10^{30}$ GeV (at $T=T_c$).

Of course, for values of $M_h$ and $M_t$ that bring the SM potential closer to being stable  (although such values are experimentally disfavored) the critical $T_c$ could be much smaller.
In principle it is even possible to make $T_c$ lower that the instability scale  $\Lambda_I$ if the non-standard minimum at high field values is sufficiently shallow. (In fact, when the value of the potential 
at the non-EW minimum goes to zero, also $T_c\ra 0$).

An alternative possibility for using a thermal probe of the potential instability arises from the realization that thermal fluctuations in the early universe can also trigger the decay of the EW vacuum \cite{Tdecay,cosmostab0,urbano,SSTU}. This process is the thermal analogue of the vacuum decay by quantum fluctuations and also proceeds via nucleation of a critical  bubble
with the decay rate $\Gamma$ controlled by an $O(3)$ symmetric bounce solution of energy $E_3(T)$, with $\Gamma \sim e^{-E_3(T)/T}$. With the use of the Nielsen identity at finite $T$ one can then prove that $E_3(T)$ is $\xi$-invariant in a way that parallels the proof for the $T=0$ decay. 
One can then ask what is the maximal temperature that our EW vacuum can survive without decaying (and this can be used to set an upper bound to the reheating temperature, see \cite{cosmostab0}). However, for the central values of $M_h$ and $M_t$ thermal effects stabilize the potential 
without destabilizing the EW vacuum and no such limit can be set
\cite{stab2.5,urbano}.

%%%%%%%%%%%%%%%%%%%%%%%%%%%%%%%%%%%%%%%%%%%%%%%%%%%
 \section{Probing the Instability Scale with Inflation\label{sec:HI}}
%%%%%%%%%%%%%%%%%%%%%%%%%%%%%%%%%%%%%%%%%%%%%%%%%%%

 Another way to characterize  the  instability scale of the SM Higgs potential in a  $\xi$-independent way is to consider the   SM in a de Sitter background  with a constant Hubble rate $H_I$. This  de Sitter stage  can be the one taking place  at primordial epochs, namely during the inflationary stage (believed to solve the main problems of standard cosmology), or it can be regarded as a purely fictitious one 
with the goal of defining a   $\xi$-independent instability scale. 

We assume that the Higgs is  minimally coupled to gravity and hence is effectively massless during inflation. (In a realistic inflationary stage, we assume no direct coupling of the Higgs to the inflaton field).
Under these circumstances,  the Higgs 
 field develops fluctuations with amplitude proportional to $H_I$  \cite{cosmostab0,cosmostab,Zurek}. Large amplitude fluctuations of the Higgs field during
the de Sitter phase  are dangerous as the SM Higgs can fluctuate beyond the instability region 
 of the effective potential. The higher is the
 Hubble rate, the larger is the probability for this phenomenon to happen, thus allowing us to define a value of the  Hubble rate  as a $\xi$-independent measure of the SM instability scale.
 
The fluctuations of the Higgs are governed by  a Langevin-like  equation obtained by the following procedure. One takes the equation of motion (in an inflationary background) for the Higgs field and splits the latter in short  and  long wavelengths, where the separation length scale is roughly the Hubble radius. Integrating out the short modes, one obtains a Langevin equation \cite{Staro,TW}
for the long mode field (that we keep calling $h$) in which the effect of the short modes is to generate a white noise $\eta$ sourcing the long mode fluctuations. More precisely, the time derivative of the original field is split into $dh/dt-\eta$.
Since one is interested in wavelengths larger than the Hubble radius, one can neglect the gradients of the long modes (and therefore truncate the starting equation at the lowest order in derivatives)
obtaining the Langevin equation:
\be
\hbox{\sc Langevin}[h]\equiv \sqrt{Z}\left( \frac{dh}{dt}-\eta\right)+\frac{1}{3H_I \sqrt{Z}}V'=0\ ,
\label{Langevin}
\ee
where $Z=Z(h)$ is the function that multiplies the kinetic term of $h$ in the effective action expanded in derivatives, as in Eq~(\ref{Sder}). To arrive at Eq.~(\ref{Langevin}), terms which are second order in derivatives, like  $(dh/dt)^2$ or $d^2h/dt^2$, are neglected. The two-point correlation function for the  noise term satisfies
\be
\langle \eta(t)\eta(t')\rangle = \frac{H_I^3}{4\pi^2 Z}\delta(t-t')\ .
\label{etaeta}
\ee
Since we previously showed that the equation of motion for the Higgs field is  $\xi$-independent and the Langevin equation stems from the  equation of motion by a simple splitting of modes,  the Langevin equation turns out to be $\xi$-independent:
\be
\xi \frac{d}{d\xi} \hbox{\sc Langevin}[h] =0\ .
\ee
This can be checked explicitly using the derivative expansion of the Nielsen identity (see Appendix~\ref{app:derexp}), paying attention to the fact that we also have to split in long and short modes  the equation, $\xi dh/d\xi=C(h)$, that controls the implicit $\xi$-dependence of the solutions of the equations of motion. This splitting leads to\footnote{Note also that the result for $\xi d\eta/d\xi$ in (\ref{splitdhdxi}) is
consistent with the correlator in (\ref{etaeta}) and the $\xi$-dependence of $Z$ to the same order of approximation (see Appendix~\ref{app:derexp}).}
\be
\xi\frac{dh_L}{d\xi}= C(h_L)\ ,\quad\quad 
\xi \frac{d\eta}{d\xi} =  C'(h_L)\eta\ .
\label{splitdhdxi}
\ee
In addition, the $\xi$-derivative of the Langevin equation leads to  terms (proportional to the $D$ and $\tilde D$ functions of the derivative expansion in  Appendix~\ref{app:derexp}) that contain two powers of potential field derivatives. Such terms are of order higher than the linear order kept for the Langevin equation, and can  be neglected.

The first equation in (\ref{splitdhdxi}) implies \cite{cosmostab2} that, if we have a solution $h_L(\xi)$ of the Langevin  equation for a given value of $\xi$, we automatically obtain a solution for $\xi+ d\xi$ by the shift $h_L(\xi)+ C[h_L(\xi)]  d\xi /\xi$.
This  is true also because the other parameter entering the Langevin equation, namely the Hubble rate $H_I$, is $\xi$-independent. Indeed, $H_I$ is determined by Einstein's equations and the $00$-component of the energy momentum tensor, both of which are $\xi$-independent objects.

\begin{figure}[t]
\begin{center}
  \includegraphics[width=0.8\textwidth]{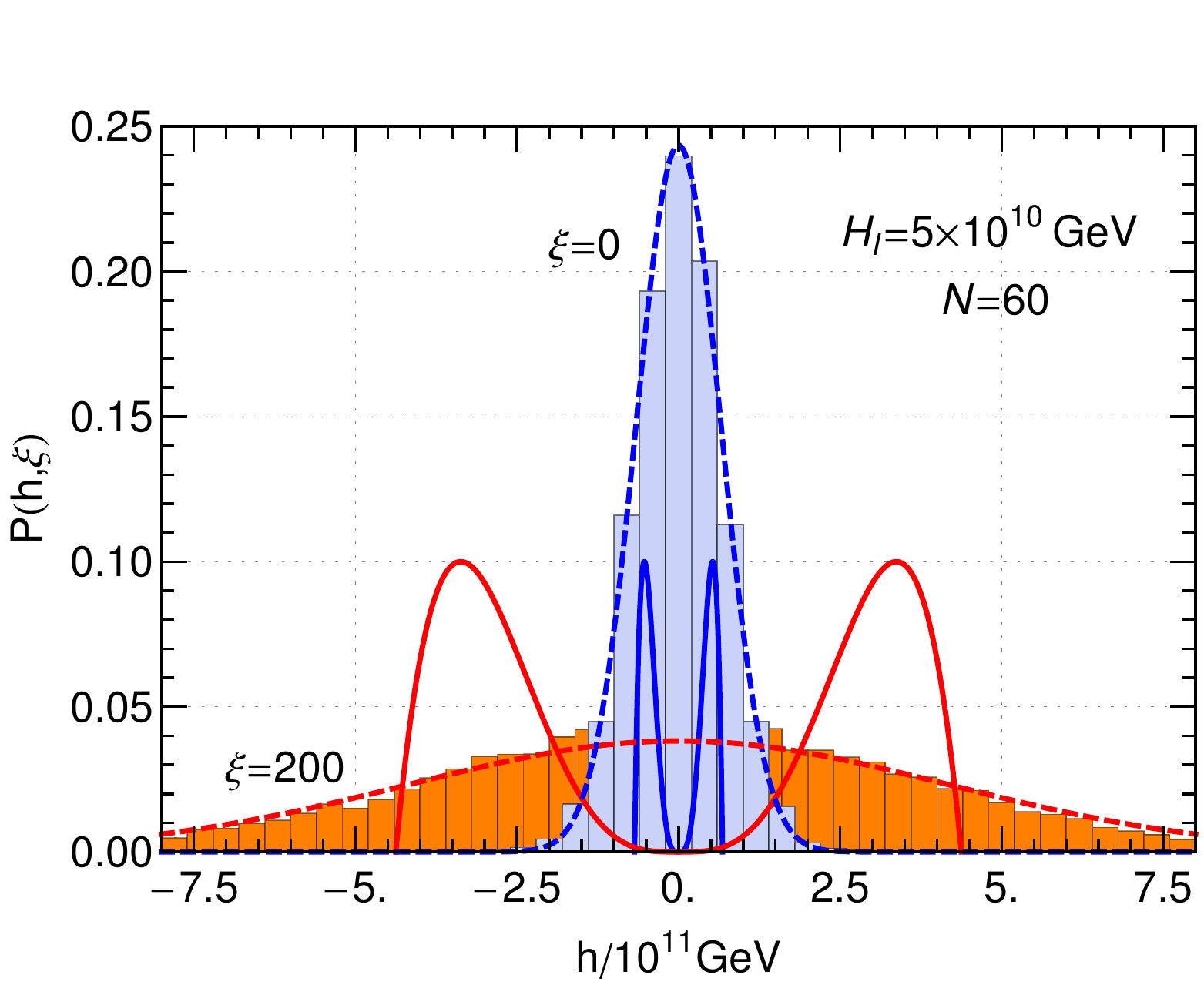}
\end{center}
\caption{\label{fig:HiggsDist}%
\em Illustration of the $\xi$-dependence of the 
fluctuating Higgs field  during an inflationary period with $N=60$ efolds and $H_I=5\times 10^{10}$ GeV and  $\xi=0$ and $200$. The histograms correspond to $10^4$ runs of the Langevin equation and the probability distribution function shown (dashed lines) is a Gaussian approximation of width $\sqrt{N} H_I/(2\pi\sqrt{Z})$, which describes well the numerical result of the Langevin runs. For comparison, the corresponding Higgs potentials are shown by the solid lines. The integrated probability between the potential maxima, $\sim 0.6$ in this example, is $\xi$-independent.
}
\end{figure}

\begin{figure}[!t]
\begin{center}
  \includegraphics[width=0.45\textwidth]{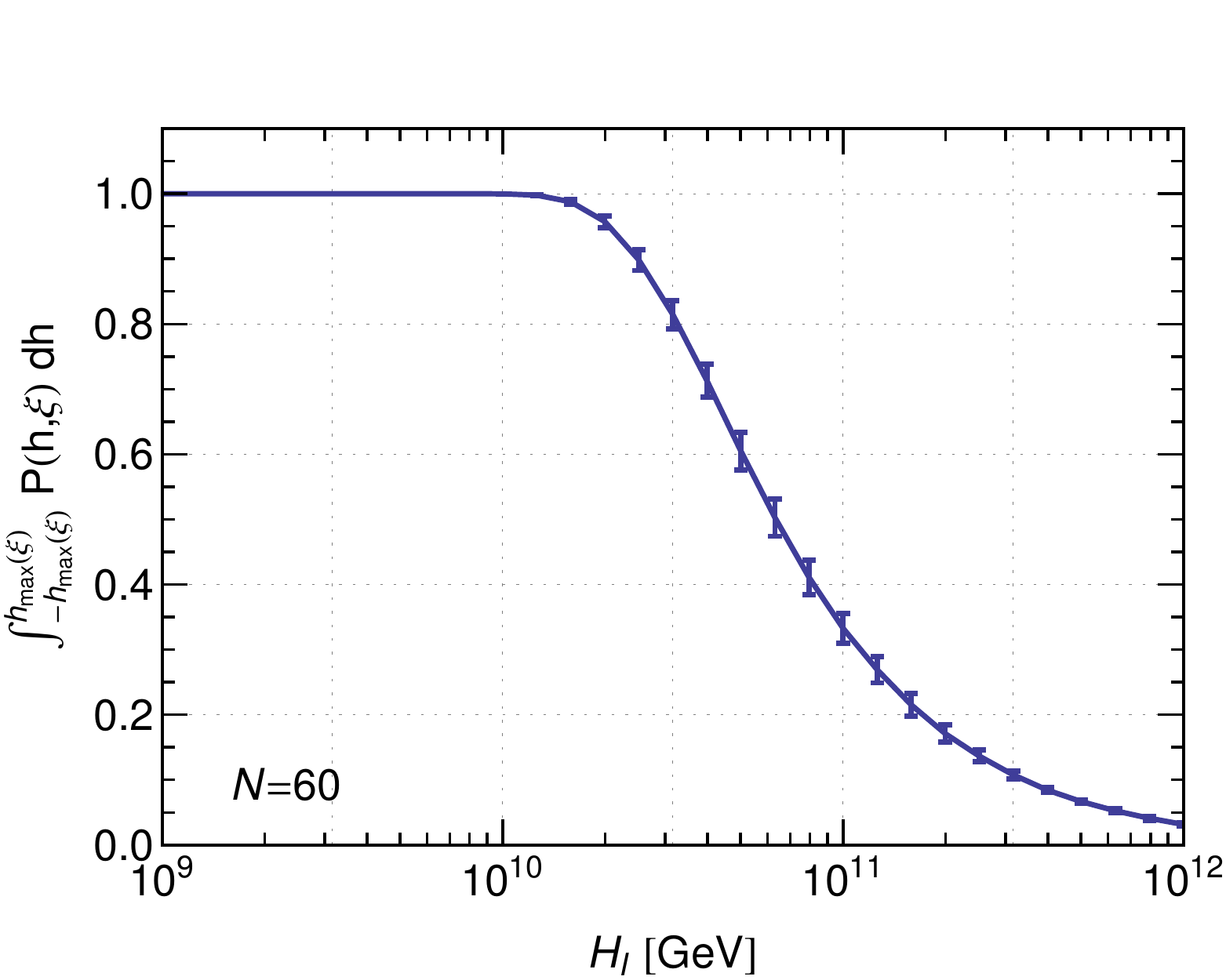}
   \includegraphics[width=0.45\textwidth]{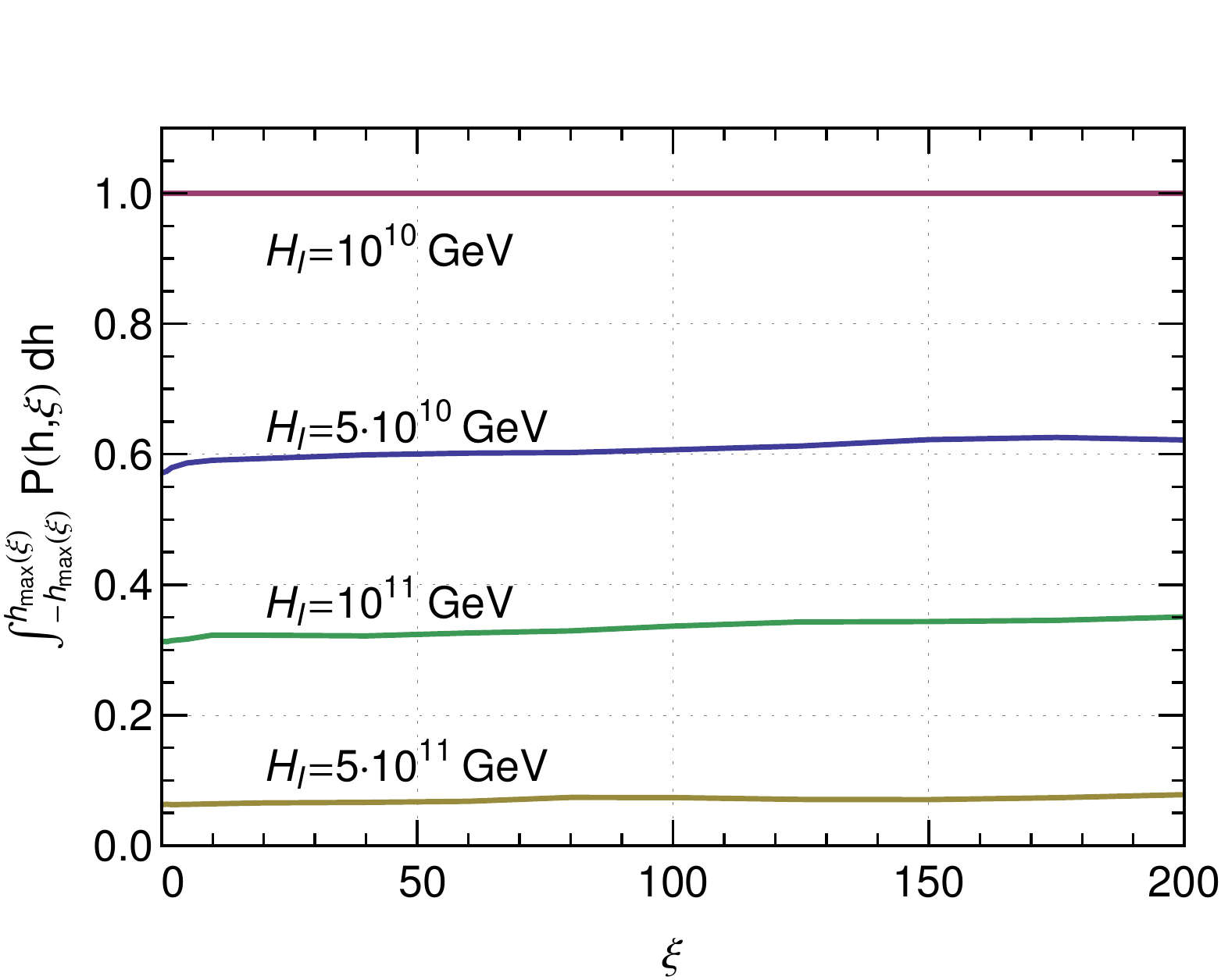}
\end{center}
\caption{\label{fig:Pxi}%
\em Integrated probability for the Higgs to remain within the potential maxima, $\int_{-h_{max(\xi)}}^{h_{max}(\xi)}P(h,\xi)dh$
for a period of $N=60$ e-folds of inflation, as a function of the Hubble rate $H_I$ (left plot)
and as a function of $\xi$ for several values of $H_I$ (right plot). In the left plot, the error bars show the  residual gauge-fixing  uncertainty when varying $\xi=(0,200)$.}
\end{figure}

Alternatively, instead of using many times the stochastic Langevin equation to sample the behaviour of the Higgs field, one can define a probability density function ${\cal P}(h,t)$, so that the probability of
finding at a given time $t$ the Higgs field in the interval $(h,h+dh)$
is  ${\cal P}(h,t)dh$. The function ${\cal P}(h,t)$ is obtained as a solution
to the  Fokker-Planck equation 
\be\label{FP}
\hbox{\sc FokkerPlanck}[{\cal P}(h,t)]\equiv \frac{1}{\sqrt{Z}}\frac{\partial}{\partial h}\left\{\frac{1}{\sqrt{Z}}\left[
\frac{\partial}{\partial h}\left(\frac{H_I^3}{8\pi^2}\frac{\cal P}{\sqrt{Z}}\right)+\frac{1}{3H_I} \frac{{\cal P}V' }{\sqrt{Z}}
\right]\right\}-\frac{1}{\sqrt{Z}}\frac{\partial {\cal P}}{\partial t}=0\ ,
\ee
which follows directly\footnote{To obtain (\ref{FP}) from (\ref{Langevin}), one must use Stratonovich's prescription rather than It\^o's, see \cite{Risken} for details.}
from the Langevin equation (\ref{Langevin}).  As this equation describes the same physics as the Langevin equation, it is also 
$\xi$-invariant:
\be
\frac{\partial}{\partial \xi }\hbox{\sc FokkerPlanck}[{\cal P}(h,t)]=0\ ,
\ee
where we use only a partial derivative because in the Fokker-Planck equation $h$ is a dummy variable, without any implicit dependence
on $\xi$.
To show this $\xi$ independence explicitly (and the previous remarks for the Langevin equation also apply here) one needs to know how ${\cal P}(h,t)$ changes when $\xi\ra \xi +d\xi$. This can be calculated as follows: for an arbitrary function $F(h)$ we 
can define its average as
\be
\langle F(h)\rangle =\int F(h') {\cal P}(h',t,\xi) dh'\ .
\ee 
To find how this average depends on $\xi$ remember that the Langevin solutions change as $h_L\ra h_L + C(h_L)d\log\xi$ and therefore, for $\xi+d\xi$ the equation above reads
\be
\langle F(h)\rangle + \langle F'(h)C(h)\rangle d\log\xi
= \int F(h') \left[{\cal P}(h',t,\xi) +\xi\frac{\partial{\cal P}(h',t,\xi)}{\partial \xi} d\log\xi\right]dh'\ ,
\ee 
from which, after integration by parts in the LHS integral we get\footnote{The same result can be obtained noting that the probability ${\cal P}(h,\xi)dh$ must equal the probability ${\cal P}(h+C(h)d\log\xi,\xi+d\xi)d(h+C(h)d\log\xi)$.}
\be
\xi \frac{\partial{\cal P}}{\partial \xi} =-\frac{\partial}{\partial h}[{\cal P} C]\ .
\label{Pxi}
\ee
We see that, contrary to what happens with the effective potential, ${\cal P}$ is not $\xi$-independent. However, the integrated probability, defined as
\be
P(h,t)= \int^h_{-\infty}  {\cal P}(h',t) dh'\ ,
\ee
is $\xi$-independent, provided the field interval of integration is 
changed according to the usual rule $\xi dh/d\xi=C(h)$. This implies, in particular, that the probability of finding the Higgs field after a given time within
the interval $(-h_{\rm max},h_{\rm max})$, where $h_{\rm max}$ is the field value corresponding to the potential maximum of the barrier separating the EW vacuum from the instability region)  is $\xi$-independent \cite{cosmostab2}, even though the value of $h_{\rm max}$ itself depends on $\xi$. Figure~\ref{fig:HiggsDist} shows the Higgs potential (solid lines) for two values of $\xi=0,200$ and the Higgs probability distribution (dashed lines), after $N=60$ efolds with $H_I=5\times 10^{10}$ GeV with the Higgs field starting at the origin. Both the potential and the Higgs probability distribution change with $\xi$ but we have checked that 
the integrated probability in $(-h_{\rm max},h_{\rm max})$ is $\xi$-independent with $\sim 9\%$ precision. Figure \ref{fig:Pxi} illustrates this $\xi$-independence of the integrated probability.

This result allows to define a $\xi$-independent measure of the instability scale based on the Hubble rate of inflation, $H_I$. One can, for instance, determine the value of $H_I$ for which the  integrated probability between two  Higgs values is larger (or smaller) than some critical number within a given amount of time. One can choose for instance the critical probability to be larger than $e^{-3 N}$,  ($N$ being the number of $e$-folds), so that it is likely to find the Higgs outside the safe interval $(-h_{\rm max},h_{\rm max})$ in any of the $\sim e^{3N}$ causally independent regions that are formed during the de Sitter stage. One gets \cite{cosmostab2} that, for $N=60$ and $h_{\rm max}\simeq 5\times 10^{10}$ GeV (the value calculated in Landau gauge for the central values of $M_h$ and $M_t$),  this happens if $H_I\geq 0.2\times  10^{10}$ GeV, a scale that is quite close to the naive instability scale calculated in Landau gauge\footnote{ Note that the bound is quite insensitive to the choice $e^{-3 N}$ of the critical probability. If one uses instead an ${\cal O}(1)$ number the bound changes by a factor $\sim \sqrt{3N}\sim {\cal O}(10)$ to $H_I\geq 2\times  10^{10}$ GeV. Using $e^{-3 N}$ corresponds to the most conservative bound,  and ${\cal O}(1)$ to the less conservative one. A more precise evaluation would require a detailed knowledge about the spatial distribution of the Higgs fluctuations with $|h|>h_{\rm max}$ at the end of inflation. We stress that this uncertainty is unrelated to the gauge dependence. The bound is gauge fixing independent for any choice of the
critical probability.}.

We close this section with a final remark on the use of the Fokker-Planck equation in this context. In order to use this equation
to describe the fluctuations of the Higgs field during inflation 
in the SM unstable potential, it is necessary to use suitable boundary conditions to describe the probability leakage towards the instability region. In practice this is done by identifying the point $h_*$ beyond which the fluctuating field rolls down to the instability and never moves back towards the stable region and setting ${\cal P}(h_*)=0$ there. The original discussion of this problem \cite{cosmostab0} used the barrier maximum, $h_*=h_{\rm max}$, for this purpose while a more correct  analysis \cite{Zurek} identified $h_*$ with the field value beyond $h_{\rm max}$ at which quantum fluctuations uphill [controlled by the term $\propto H_I^3$ in (\ref{FP})] can no longer beat the classical rolling downhill [controlled by the term $\propto V'$ in (\ref{FP})]. This analysis was  further validated by using the Langevin approach in \cite{cosmostab2}.
The field value $h_*$ corresponds to the point at which the two terms in the Fokker-Planck equation compensate each other giving $\partial {\cal P}/\partial t=0$ and can be roughly estimated as the value for which the quantum jumps $(\Delta h)_q\sim H_I/(2\pi \sqrt{Z})$ in a time interval $\Delta t\sim 1/H_I$  equal the classical displacement
$(\Delta h)_{cl}\sim V'/(3H_I^2 Z)$, leading to the condition
\be
\frac{V'}{Z}\simeq \frac{3H_I^3}{2\pi}\ ,
\label{hstar}
\ee
to determine $h_*$. As one would expect, $h_*$ is a gauge-dependent quantity but it can be shown, using the leading $\xi$-dependence of $V$ and $Z$  that the condition (\ref{hstar}) is $\xi$-independent if $h_*$ is transformed in the usual way, with $\xi dh_*/d\xi=C(h_*)$. Formally one can also check that the Fokker-Planck equation at $h_*$, using further that ${\cal P}(h_*)=0$,
and $\partial {\cal P}/\partial t=0$, leads to the condition
\be
\frac{H_I^3}{8\pi^2}\left[{\cal P}''-\frac{3Z'{\cal P}'}{2Z}\right]=
-\frac{V'{\cal P}'}{3H_I}\ .
\label{quantclass}
\ee
which can also be shown to be $\xi$-invariant, if one further uses the $\xi$-dependence of ${\cal P}$ obtained in (\ref{Pxi}).

%%%%%%%%%%%%%%%%%%%%%%%%%%%%%%%%%%%%%%%%%%%%%%%%%%%
\section{Conclusions}
%%%%%%%%%%%%%%%%%%%%%%%%%%%%%%%%%%%%%%%%%%%%%%%%%%%

It has been known since a long time
that the SM might develop an instability at large values of the Higgs field, possibly signaling the appearance of new physics in that range of energies, if the Higgs mass turned out to be low enough. With the discovery of the Higgs at the LHC, and the theoretical refinements needed for the stability bound calculations, this possibility turned out to be realized, with the Higgs mass very close (but below) the value needed for stability.

The naive definition of the instability scale as the value of the Higgs field at which the Higgs effective potential drops below the value of the electroweak minimum ($\sim 10^{10}$ GeV in Landau gauge) is not physical as it depends on the choice of the gauge-fixing and results in an uncertainty of two orders of magnitude. While such gauge dependence has also been known for a long time, it has now become more relevant and has attracted some attention in recent literature.

Current direct and indirect experimental probes do not show any evidence of  new physics beyond the SM and it becomes more and more likely that 
the latter might be valid up to energy scales much higher than the TeV, maybe up to the Planck scale. It is therefore timely  to provide
gauge-invariant and physical descriptions of the scales associated to the SM vacuum instability.

In this paper we have proposed several ways to characterize in a physical way such scales. In particular, we have 
shown that  the mass scale $\Lambda$  required to stabilize the effective potential through non-renormalizable operators is a  gauge-independent
quantity. We have shown that for non-renormalizable operators of high order, the scale $\Lambda$ is close to the naive instability scale calculated in Landau gauge. We have also demonstrated the gauge-invariance of three other scales:  the inverse of the  critical radius of the critical bubbles that mediate vacuum decay, the critical temperature at which our  electroweak  minimum and the one at very large Higgs  field values become degenerate and finally the scale associated to the Higgs instability during inflation. Being all these scales gauge-invariant, they allow to draw 
physical conclusions on the associated physics, for instance on how new physics    influences   the EW vacuum lifetime.

Although we have focused on the SM potential instability, which offers the main motivation for this work, our results are of wider 
relevance and can be applied to other models (with gauge degrees of freedom) that feature similar instabilities and/or several potential vacua.

%%%%%%%%%%%%%%%%%%%%%%%%%%%%%%%%%%%%%%%%%%%%%%%%%%%
\section*{Acknowledgments}
%%%%%%%%%%%%%%%%%%%%%%%%%%%%%%%%%%%%%%%%%%%%%%%%%%%

We thank Kaladi S. Babu, Pepe Barb\'on, Enrico Bertuzzo, Joan Elias-Mir\'o, Dani Figueroa, Jeff Fortin, Luca di Luzio, Werner Porod, Alessandro Strumia, Nikos Tetradis, James Unwin and especially Gian Giudice for very useful discussions. J.R.E. thanks the CERN TH-Division for hospitality and partial financial support during several stages of this project.
This work has been partly supported by the ERC
grant 669668 -- NEO-NAT -- ERC-AdG-2014, the Spanish Ministry MINECO under grants  FPA2013-44773-P and
FPA2014-55613-P, the Severo Ochoa excellence program of MINECO (grant SEV-2012-0234) and by the Generalitat grant 2014-SGR-1450. MG and TK acknowledge partial support by the Munich Institute for Astro- and Particle Physics (MIAPP)
of the DFG cluster of excellence  ``Origin and Structure of the Universe''. We also acknowledge support by the German Science Foundation (DFG) within the Collaborative Research Center (SFB) 676 ‘Particles, Strings and the Early Universe. A.R. is supported by the Swiss National Science Foundation (SNSF), project {\sl Investigating the
Nature of Dark Matter}, project number: 200020-159223.

\appendix
\numberwithin{equation}{section}

%%%%%%%%%%%%%%%%%%%%%%%%%%%%%%
\section{RGES for Higher-Dimensional Operators\label{app:RGES}}
%%%%%%%%%%%%%%%%%%%%%%%%%%%%%%

Consider a tree-level Higgs potential of the form
\be
  V_0(h) = -\frac12 m^2 h^2 + \frac{\lambda}{4}h^4 +  \sum_n\frac{c_nh^n}{2^{n/2}\Lambda^{n-4}}\,,
\ee
where summation runs over $n=6,8,\dots$.
The radiatively corrected potential  is independent of the renormalization scale, as described by the Callan-Symanzik equation 
\be
\frac{d V}{d\mu}=\left(\mu\frac{\partial}{\partial\mu} + \sum_i\beta_{\lambda_i}
\frac{\partial}{\partial\lambda_i}+\gamma h \frac{\partial}{\partial h }\right)V=0\ ,
\ee
where the $\lambda_i$'s represent all couplings and mass parameters entering the potential (including the  gauge-fixing $\xi$ parameters) and $\gamma$ is the Higgs anomalous dimension ($\gamma\equiv d\log h/d\log\mu$). The one-loop RG equation for $c_n$ can be obtained from this equation via the explicit $\mu$-dependence of the one-loop effective potential and knowledge of the anomalous dimension $\gamma$, which at one loop is not affected by the irrelevant operators. In Fermi gauge we obtain
\be
 \sum_n\frac{\partial V_0}{\partial c_n}\beta_{c_n} = \frac{\kappa}{2}\left[M_H^4(h) + 3M_G^4(h) \right] + \gamma_r h V_0'(h)\,,
 \label{dVdmu}
\ee
where the field-dependent Higgs and Golstone masses are given in Eq.~(\ref{MHMG}) and $\gamma_r\equiv \gamma - \kappa (\xi_B {g'}^2 + 3\xi_W g^2)/4$ in the Standard Model.  Using the one-loop
anomalous dimension in Fermi gauge \cite{DiLuzio} one finds for the Standard Model
\be
  \gamma_r = \kappa\left(3y_t^2-\frac34 {g'}^2-\frac94 g^2\right)\,.
  \label{gammar}
\ee
Note that the $\xi$-dependence cancels in $\gamma_r$,
such that the beta functions are gauge-independent, as expected.
Inserting (\ref{gammar}) and the potential $V_0(h)$ in Eq.~(\ref{dVdmu}) and Taylor expanding in $h$,  one gets the one-loop beta function given in \eqref{eq:betan} by matching same powers of $h$ in both sides of the equation (and taking the limit $m^2\to 0$).

%%%%%%%%%%%%%%%%%%%%%%%%%%%%%%%%%%%%%%%%%%%%%%%%%%%
\section{Gauge Dependence and Derivative Expansion\label{app:derexp}}
%%%%%%%%%%%%%%%%%%%%%%%%%%%%%%%%%%%%%%%%%%%%%%%%%%%

In this Appendix we assume that a derivative expansion of the action is applicable and discuss how the gauge parameter (generically denoted as $\xi$) enters the action, the equation of motion, and the energy-momentum tensor up to higher order
corrections in the expansion. We give explicit results up to fourth
order in derivatives for the case of a single scalar field $ \phi$, therefore extending previous work done to ${\cal O}(\partial^2)$ \cite{GK,Metaxas}.

We start from the expanded effective action
\bea
S &=& \int d^4 x \left[\frac12 Z( h) (\partial \phi)^2 - V(\phi)\right.\nonumber\\
&&+\left.\frac14 Z_2(\phi)  (\partial \phi)^4 + \frac12 Z_3(\phi) 
(\partial \phi)^2(\partial^2 \phi) + \frac12 Z_4(\phi) (\partial^2 \phi)^2 +  {\cal O}(\partial^6)
\right]\ .
\label{S4}
\eea
There are only three $Z_i(\phi)$ functions at fourth order 
in derivatives as other possible terms  can be eliminated integrating by parts. The $\xi$-dependence of this action is dictated by the Nielsen identity that we write as
\be
\xi\frac{\partial S}{\partial \xi} + \int d^4x K[\phi(x)] \frac{\delta S}{\delta \phi}=0\ .
\ee
The functional $K[\phi(x)]$ can also be expanded as
\be
K[\phi] = C(\phi) - D(\phi)(\partial \phi)^2 - \tilde D(\phi) (\partial^2 \phi) +  {\cal O}(\partial^4)\ ,
\label{K2}
\ee
and, for our purposes below, it is enough to keep up to second order in derivatives only.
The explicit form of the equation of motion  for $\phi$, is 
\be
\frac{\delta S}{\delta \phi}  \equiv {\rm EoM}[\phi] =0\ ,
\nonumber
\ee
with
\bea
&{\rm EoM}[\phi]=& -V' +\frac12 Z' \dpsq - \partial_\mu[Z\partial^\mu\phi] +\frac14
Z_2' \dpf -\partial_\mu[Z_2\partial^\mu\phi\dpsq]+\frac12 Z_3'\dsqp \dpsq \nonumber\\
&&-\partial_\mu[Z_3\partial^\mu\phi \dsqp]+\frac12\partial^2[Z_3\dpsq]+\frac12 Z_4'\dsqp^2+\partial^2[Z_4\dsqp]+{\cal O}(\partial^6)\ ,
\label{EoM4}
\eea
where primes denote field derivatives, $V'=\partial V/\partial\phi$, etc. This shows explicitly that, evaluated on solutions of the EoM, $V'$ counts as being ${\cal O}(\partial^2)$.

Plugging all the previous expansions in the Nielsen identity, which is satisfied for generic $\phi(x)$, one can derive the following $\xi$-dependence
of the functions appearing in the action (\ref{S4}):
\bea
\xi\frac{\partial V}{\partial\xi} &=& - CV'\ ,\nonumber\\
\xi\frac{\partial Z}{\partial\xi} &=&- C Z' -2 C'Z-2DV'+2(\tilde D V')'\ ,\nonumber\\
\xi\frac{\partial Z_2}{\partial\xi} &=&- C Z_2' -4 C'Z_2-2C''Z_3 -2DZ'+...\ ,\nonumber\\
\xi\frac{\partial Z_3}{\partial\xi} &=&- C Z_3' -3 C'Z_3-2C''Z_4 -2DZ'-\tilde D Z'+...\ ,\nonumber\\
\xi\frac{\partial Z_4}{\partial\xi} &=&- C Z_4' -2 C'Z_4 -2\tilde DZ'+...
\label{Sxi}
\eea
The terms neglected in the last three equations involve 
$V'$ and fourth-derivative terms in the expansion (\ref{K2}) of $K[\phi]$: they contribute only at ${\cal O}(\partial^6)$ in our discussions below.

Under a change $\xi \ra \xi +d\xi$, all the functions in the action change according to the equations (\ref{Sxi}) and therefore the EoM in (\ref{EoM4}) is also modified. If $\bar\phi(\xi)$ is a solution
of the original EoM, then a solution of the EoM for $\xi \ra \xi +d\xi$ is $\bar\phi(\xi+d\xi)=\bar\phi(\xi)+d\bar\phi$
with
\be
\xi\frac{d\bar\phi}{d\xi} = K[\bar\phi(x)]\ .
\label{dphi}
\ee
This fact is obvious from the invariance of the action itself, as
the dependence above gives $\xi dS/d\xi=0$, and is one clear example of the power of using an action to discuss the symmetries in a physical problem. The explicit check 
using the EoM expanded to fourth order in the derivative expansion, 
as given in Eq.~(\ref{EoM4}), is more involved, but straightforward using (\ref{Sxi}) and (\ref{dphi}). In performing this check, one needs to evaluate the $\xi$-dependence of field-derivatives of the functions $V, Z$ and $Z_i$. This can be done simply by taking field-derivatives of Eqs.~(\ref{Sxi}), which are identities that hold for generic $\phi(x)$.  
For instance, one gets
\be
\xi\frac{\partial V'}{\partial\xi} = - C'V'- CV''\ ,
\ee
and so on. In this way one is able to eliminate all $\xi$ derivatives.
In order to complete the check one must also get rid of field derivatives of the potential. The first derivative, $V'$, is eliminated  by using the EoM (\ref{EoM4}). Higher order derivatives like $V''$ and $V'''$ can then be eliminated by taking spacetime derivatives of the EoM (which holds at every spacetime point). In this way one can get rid of the combinations $\partial_\mu V'=V''\partial_\mu\bar\phi$ and $\partial^2 V'=V''\partial^2\bar\phi+V'''(\partial\bar\phi)^2$, etc. After carrying through this program one gets
\be
\frac{d}{d\xi}{\rm EoM}[\bar\phi] =0\ ,
\label{dEoMdxi}
\ee
to order ${\cal O}(\partial^6)$. This completes the proof
presented in \cite{cosmostab2} to ${\cal O}(\partial^4)$ with one important improvement: we have shown now that no assumption about the subleading role of the functions $D(\phi)$ and $\tilde D(\phi)$ is needed. Note also that,
to order  ${\cal O}(\partial^4)$, it is enough to use $\xi d\bar\phi/d\xi=C(\bar\phi)$.

One can also check explicitly, in the derivative expansion approximation, that the energy-momentum tensor for solutions of the EoM is $\xi$-independent. Consider an action of the form
\be
  S = \int d^4x \sqrt{-g} {\cal L}(\phi, g^{\mu\nu}\partial_\mu\phi\partial_\nu\phi, \Box \phi)\,,
\ee
where
\be
  \Box\phi = D_\mu D^\mu \phi = \frac{1}{\sqrt{-g}}\partial_\mu(\sqrt{-g}\partial^\mu\phi )= \frac{1}{\sqrt{-g}}\partial_\mu(\sqrt{-g}g^{\mu\nu}\partial_\nu\phi)\,,
\ee
is the covariant d'Alembert operator applied to a scalar field. We are ultimately interested in a flat background, but keep the metric in order
to derive the energy momentum tensor. The equation of motion will only be needed on a Minkowski background, where it reads
\be
  {\cal L}_1 - \partial^\mu\left(2{\cal L}_2 \partial_\mu\phi\right) + \Box {\cal L}_3 = 0\,,
\ee
where we used the short-hand notation ${\cal L}_i=\partial{\cal L}(X_1,X_2,X_3)/\partial X_i$.
To derive the energy-momentum tensor we use the standard relations
\bea
  \frac{\partial}{\partial g_{\rho\sigma}(y)} g^{\mu\nu}(x) &=& -\frac12 (g^{\mu \rho}g^{\nu \sigma}+g^{\mu \sigma}g^{\nu \rho})\delta(x-y)\,,\\
  \frac{\partial}{\partial g_{\rho\sigma}(y)} (\sqrt{-g})^n &=& \frac{n}{2}(\sqrt{-g})^n g^{\rho\sigma}\delta(x-y)\,.
\eea
This gives, on a Minkowski background,
\be
  \frac{\partial}{\partial g_{\rho\sigma}(y)}\Box\phi(x) = -2\delta(x-y)\partial^\rho\partial^\sigma\phi-\partial^\rho\delta(x-y)\partial^\sigma\phi-\partial^\sigma\delta(x-y)\partial^\rho\phi
+ g^{\rho\sigma}\partial_\eta\delta(x-y)\partial^\eta\phi \;.
\ee
Using these relations one obtains
\be
  T_{\mu\nu}[\phi] = -\frac{2}{\sqrt{-g}}\frac{\delta S}{\delta g_{\mu\nu}}  =2\partial_\mu\phi\partial_\nu\phi\,{\cal L}_2
  -\partial_\mu{\cal L}_3 \partial_\nu\phi - \partial_\nu{\cal L}_3 \partial_\mu\phi -g_{\mu\nu}\left[{\cal L}-\partial_\eta({\cal L}_3\partial^\eta\phi)\right]\;.
\ee
Using the equation of motion one can check that $\partial^\mu T_{\mu\nu}=0$. For the action (\ref{S4}), one gets
\begin{align}
T^{\mu\nu} &= g^{\mu\nu}\left[V-\frac{Z}{2}\dpsq-\frac{Z_2}{4}\dpf+
Z_3\partial_\rho\phi\ \partial_\sigma\phi\ \partial^{\rho\sigma}\ph
+
Z_4\partial_\rho\phi\ \partial^\rho\partial^2\phi
+Z'_4\dsqp\dpsq
\right.\nonumber\\
&\left.
+\frac{Z_4}{2} \dsqp^2
+\frac{Z'_3}{2}\dpf
\right]
-Z_3\left(\partial^\mu\phi\ \partial^{\nu\rho}\phi+\partial^\nu\phi\ \partial^{\mu\rho}\phi\right) \partial_\rho\phi
-Z_4(\partial^\mu\phi\ \partial^\nu+\partial^\nu\phi\ \partial^\mu)\partial^2\phi \nonumber\\
&-Z_4(\partial^\mu\phi\ \partial^\nu+\partial^\nu\phi\ \partial^\mu)\partial^2\phi +\partial^\mu\phi\ \partial^\nu\phi
\left[Z+(Z_2-Z_3')\dpsq+(Z_3-2Z'_4)\dsqp
\right] +{\cal O}(\partial^6) .
\label{Tmunu4}
\end{align}
Using the same relations discussed above regarding the EoM, and in particular, the $\xi$-dependence of the solutions of the EoM given by (\ref{dphi}), one is able to prove that 
\be
\frac{d}{d\xi}T^{\mu\nu}[\bar\phi]=0\ ,
\label{dTmunudxi}
\ee
to order ${\cal O}(\partial^6)$,
so that the energy and momentum densities are $\xi$-independent quantities.

The discussion above has been performed for the Minkowski
action and EoM but, for applications to tunneling rates and bounce solutions one uses instead the Euclidean action and EoM. However, the good $\xi$-independence properties carry over to the Euclidean
case. To see this explicitly, note first that in the Minkowskian proofs of $\xi$ independence presented above we never had to deal with the exact form of  the metric.  However, the Euclidean action is obtained by the replacement $t\ra -i t_E$, under which
$\dpsq \ra - (\partial_E \phi)^2$, $\dsqp \ra -\partial_E^2\phi$.
These sign flips could be assigned to a parity transformation
under which 
\be
Z\ra -Z,\quad Z_i\ra Z_i,\quad V\ra V,\quad C\ra C,\quad 
D\ra -D,\quad \tilde D\ra -\tilde D\ ,
\ee
with $i=2,3,4$.
It is immediate to see that  
Eqs.~(\ref{Sxi}), (\ref{dEoMdxi}) and (\ref{dTmunudxi}) are invariant under this parity transformation so that the $\xi$-invariance they express also holds in the Euclidean case.

%%%%%%%%%%%%%%%%%%%%%%%%%%%%%%%%%%%%%%%%%%%%%%%%%%%
\section{Nielsen Identity at Finite Temperature \label{app:NIT}}
%%%%%%%%%%%%%%%%%%%%%%%%%%%%%%%%%%%%%%%%%%%%%%%%%%%

The Nielsen identity describing the gauge dependence of the effective potential extended to finite temperature  takes the form given in Eq.~(\ref{dVdxiT}). Here we show the explicit forms of
the effective potential $V_T$ and the $C_T$ function at finite $T$ for the SM in Fermi gauge.

The calculation of $V_T$ is by now standard. At one-loop order one gets
\be
V_T(h)= V_0(h) + V_1(h,T)\ .
\ee
Here $V_0(h)$ is the tree-level potential of Eq.~(\ref{V0}).
The one-loop term $V_1(h,T)$ includes both the $T=0$
radiative corrections, as explicitly given in Eq.~(\ref{V1}), and the  free-gas approximation for the thermal effects and depends only on the $h$-dependent particle masses $M_\alpha$ and their number of degrees of freedom $N_\alpha$ (taken negative for fermions), with 
$\alpha$ labelling different particle species, see (\ref{spectrum}).

Both $T=0$ and finite $T$ contributions to the one-loop potential
$V_1(h,T)=V_1(h,0)+\Delta_T V_1(h,T)$,  arise in a compact way in the imaginary-time formalism expression
\be
V_1(h,T) = \sum_\alpha N_\alpha J_\alpha(M^2_\alpha,T)\ .
\label{VT}
\ee 
where the functions $J_\alpha$, defined as
\be
J_\alpha(M^2_\alpha,T) \equiv \frac12  \sumintK \log(K^2 + M_\alpha^2)\ ,
\label{Ja}
\ee
are the finite-$T$ generalization of the $J^0_\alpha(M^2_\alpha)$ functions introduced in Eq.~(\ref{V1}).
Inside the logarithm $K^2$ stands for the Euclidean momentum, after 
$k_0\ra i k_0$, while the integral-sum stands for
\be
\sumintK \equiv   T \sum_{n} \int \frac{d^{3-2\epsilon}k}{(2\pi)^{3-2\epsilon}}\ ,
\ee 
with
$\epsilon$ used for dimensional regularization.
For nonzero $T$, $k_0$ takes discrete values:  $k_0= 2\pi n T$ for bosons and $
k_0=(2n+1)\pi T$ for fermions, so that the $k_0$ integral is replaced by a sum with $n$ running over all integers.  
Note that for $T\ra 0$ we have $T\sum_n\ra \int dk_0/(2\pi)$
and one recovers the $T=0$ result (after Wick rotation), so that in fact the function (\ref{Ja}) contains the finite temperature piece
on top of the $T=0$ one in a single formula.
The thermal correction is
\be
\Delta_T V_1(h,T)=\sum_\alpha N_\alpha T\int d^3k 
\log\left[
1 +s_\alpha e^{-\sqrt{k^2+M_\alpha^2}}\right]\ ,
\ee
with $s_\alpha=-$ ($+$) for bosons (fermions). 
When $T\gg M_\alpha$, a high-$T$ expansion for the
potential gives
\bea
V_1(h,T) &=&\sum_{\ \alpha\, {\rm (B)}\ }N_\alpha\left[
-\frac{\pi^2 T^4}{90}+\frac{1}{24}M_\alpha^2 T^2
-\frac{1}{12\pi}M_\alpha^3 T+\frac{M_\alpha^4}{64\pi^2}\log\frac{c_B T^2}{\mu^2}+{\cal O}\left(\frac{M_\alpha^6}{T^2}\right)\right]\nonumber\\
&+&\sum_{\ \alpha\, {\rm (F)}\ }N_\alpha\left[
\frac{7\pi^2 T^4}{720}-\frac{1}{48}M_\alpha^2 T^2
+\frac{M_\alpha^4}{64\pi^2}\log\frac{c_F T^2}{\mu^2}+{\cal O}\left(\frac{M_\alpha^6}{T^2}\right)\right]\ ,
\eea
where bosonic and fermionic contributions are indicated separately by (B) and (F) respectively and $c_B=3/2+2\log(4\pi)-2\gamma_E$, $c_F=3/2+2\log\pi-2\gamma_E$ with $\gamma_E\simeq 0.577$ the Euler-Mascheroni constant. In many cases of interest (like in the study of the electroweak phase transition) a resummation of finite $T$ IR divergent terms (from $n=0$ bosonic modes) is necessary. The main effect, after resumming the so-called ring diagrams
is to replace the mass in the cubic term of the bosonic expansion above by a thermally screened one, $M_\alpha^2\ra M_\alpha^2 + {\cal O}(T^2)$, see e.g. \cite{AE} for details.

The expressions for $C_T$ in the SM (for Fermi gauge), generalize the $T=0$ ones given in Eqs.~(\ref{CB},\ref{CW}).
Going to momentum space and using the imaginary time formalism as before we arrive at the thermally corrected expressions (at one loop), 
\bea
C^{(1)}_{B,T}(h)&=&-\frac{g'}{2} \sumintK P_c(-K) P_{\chi^0,\mu}(K) K^\mu= -\frac{\xi_B}{8} {g'}^2 h\sumintK \frac{1}{(K^2+M^2_{B_+})(K^2+M^2_{B_-})}
\ , \nonumber\\
C^{(1)}_{W,T}(h)&=&-\frac{g}{2} \sumintK P_{c_a}(-K) P_{\chi^a,\mu}(K) K^\mu \\
&=&  - \frac{\xi_W}{8} {g}^2 h\sumintK \left[\frac{1}{(K^2+M^2_{B_+})(K^2+M^2_{B_-})}+\frac{2}{(K^2+M^2_{A_+})(K^2+M^2_{A_-})}
\right]\ ,\nonumber
\eea
where the intermediate expressions are written in terms of the ghost propagators and the Goldstone-gauge mixed ones and $M^2_{A_\pm}$, $M^2_{B_\pm}$ are given in Eq.~(\ref{spectrum}).

It is convenient to rewrite the previous expressions  in terms of the functions $I_\alpha(M_\alpha^2,T)$ defined by
\be
I_\alpha(M_\alpha^2,T)\equiv 2\frac{\partial}{\partial M_\alpha^2}
J_\alpha(M_\alpha^2,T)=\sumintK \frac{1}{K^2+M_\alpha^2}\ .
\label{IJ}
\ee
Then we get
\bea
C^{(1)}_{B,T}(h)&=&\frac{1}{8}\xi_B {g'}^2 h\left[\frac{I_{B_+}-I_{B_-}}{M^2_{B_+}-M^2_{B_-}}\right]
\ , \nonumber\\
C^{(1)}_{W,T}(h)&=&\frac{1}{8} \xi_W{g}^2 h
\left[\frac{I_{B_+}-I_{B_-}}{M^2_{B_+}-M^2_{B_-}}+2 
\frac{I_{A_+}-I_{A_-}}{M^2_{A_+}-M^2_{A_-}}
\right]\ ,
\eea
a straightforward generalization of the $T=0$ results of Eqs.~(\ref{CB0},\ref{CW0}).
The high-$T$ expansions of these expressions are
\bea
C^{(1)}_{B,T} &=& -\frac{\kappa}{8}\xi_B {g'}^2 h\left[\log\frac{\mu^2}{c_BT^2}+\frac32+\frac{4\pi T}{M_{B+}+M_{B_-}}+ {\cal O}\left(\frac{M^2_{B_\pm}}{T^2}\right)\right]\ , 
\\
C^{(1)}_{W,T} &=& -\frac{\kappa}{8}\xi_W g^2 h\left[2\log\frac{\mu^2}{c_BT^2}+3+\frac{4\pi T}{M_{B+}+M_{B_-}}+\frac{8\pi T}{M_{A+}+M_{A_-}} + {\cal O}\left(\frac{M^2_{B_\pm}}{T^2},\frac{M^2_{A_\pm}}{T^2}\right)\right]\ , \nonumber
\eea
which agree with the leading term result presented in \cite{GK} (translated to Fermi gauge).

Finally, noting that 
\bea
\xi_B \frac{\partial V_1(h,T)}{\partial\xi_B}  &=& \frac{\partial J_{B_+}}{\partial M^2_{B_+}}\ \xi_B\frac{\partial M^2_{B_+}}{\partial\xi_B}+ (B_+\ra B_-)\ ,\nonumber\\
\xi_W \frac{\partial V_1(h,T)}{\partial\xi_W}  &=& \frac{\partial J_{B_+}}{\partial M^2_{B_+}}\ \xi_W\frac{\partial M^2_{B_+}}{\partial\xi_W}+2\frac{\partial J_{A_+}}{\partial M^2_{A_+}}\ \xi_W\frac{\partial M^2_{A_+}}{\partial\xi_W}+ (B_+\ra B_-,A_+\ra A_-)\ ,
\nonumber
\eea
and using the expressions for $M^2_{B_\pm}$ and $M^2_{A_\pm}$
given in (\ref{spectrum}) plus the relation (\ref{IJ}) and 
$\partial V_0/\partial h=M_G^2 h$, it is straightforward to check the one-loop Nielsen identities
\bea
&&
\xi_B \frac{\partial V_1(h,T)}{\partial \xi_B} + C^{(1)}_{B,T} \frac{\partial V_0(h)}{\partial h}=0\ ,\\
&&
\xi_W \frac{\partial V_1(h,T)}{\partial \xi_W} + C^{(1)}_{W,T} \frac{\partial V_0(h)}{\partial h}=0\ .
\eea

%%%%%%%%%%%%%%%%%%%%%%%%%%%%%%%%%%%%%%

\end{document}